\documentclass[journal]{IEEEtran}

\usepackage{cite,graphicx,amsmath,amssymb}
\usepackage{subfigure}
\usepackage{citesort}
\usepackage{fancyhdr}
\usepackage{mdwmath}
\usepackage{mdwtab}
\usepackage{balance}
\usepackage{xcolor}

\newtheorem{theorem}{Theorem}
\newtheorem{corollary}{Corollary}

\makeatletter
\def\ScaleIfNeeded{%
\ifdim\Gin@nat@width>\linewidth \linewidth \else \Gin@nat@width
\fi } \makeatother

\begin{document}

\title{Cooperative Non-Orthogonal Multiple Access with Simultaneous Wireless Information and Power Transfer}

 \author{Yuanwei\ Liu,~\IEEEmembership{Student Member,~IEEE,}
  Zhiguo\ Ding,~\IEEEmembership{Member,~IEEE,}

   Maged\ Elkashlan,~\IEEEmembership{Member,~IEEE,}
    and H. Vincent\ Poor,~\IEEEmembership{Fellow,~IEEE}
\thanks{Y. Liu and M. Elkashlan are with the School of Electronic Engineering and Computer Science, Queen Mary University of London, London E1 4NS, UK. (email:\{yuanwei.liu, maged.elkashlan\}@qmul.ac.uk).}
\thanks{Z. Ding and H. V. Poor are with the Department of Electrical Engineering, Princeton University, Princeton, NJ 08544, USA. (e-mail: poor@princeton.edu). Z. Ding is also with the School of Computing and Communications, Lancaster University, LA1 4WA, UK. (e-mail: z.ding@lancaster.ac.uk).}
\thanks{ This research was supported in part by the UK EPSRC under grant
number EP/L025272/1 and the U.S. National Science Foundation under Grant EECS-1343210.}
}

\maketitle
\begin{abstract}
In this paper, the application of simultaneous wireless information and power transfer (SWIPT) to non-orthogonal multiple access (NOMA) networks in which  users  are spatially randomly located is investigated. A new cooperative SWIPT NOMA protocol is proposed, in which near NOMA users that are close to the source act as energy harvesting relays to help far NOMA users. Since the locations of users have a  significant impact on the performance, three user selection schemes based on the user distances from the base station are proposed. To characterize the performance of the proposed selection schemes,   closed-form expressions for the outage probability and system throughput are derived. These analytical results demonstrate that  the use of SWIPT will not jeopardize the diversity gain compared to the conventional NOMA. The proposed results confirm that the opportunistic use of node locations for user selection can achieve low outage probability and deliver superior throughput
in comparison to the random selection scheme.


\begin{keywords}
{N}on-orthogonal multiple access, simultaneous wireless information and power transfer, stochastic geometry, user selection
\end{keywords}
\end{abstract}

\section{Introduction}
Non-orthogonal muliple access (NOMA) is an effective solution to improve spectral efficiency and has recently received significant attention for its promising application in fifth generation (5G) networks~\cite{saito2013system}. The key idea of NOMA is to realize  multiple access (MA) in the power domain which is fundamentally different from conventional orthogonal  MA technologies (e.g., time/frequency/code division MA). The motivation behind this approach lies in the fact that NOMA can use spectrum more efficiently by opportunistically exploring users' channel conditions~\cite{ding2014pairing}. In \cite{ding2014performance}, the authors investigated the performance of a downlink NOMA scheme with randomly deployed users. An uplink NOMA transmission scheme was proposed in~\cite{al2014uplink}, and its performance was evaluated systematically. In~\cite{ding2014pairing}, the impact of user pairing was characterized by analyzing the sum rates in two NOMA systems, namely, fixed power allocation NOMA and cognitive radio inspired NOMA. In \cite{ding2014letter}, a new cooperative NOMA scheme was proposed and analyzed in terms of outage probability and diversity gain.

In addition to improving spectral efficiency which is the motivation of NOMA, another key objective of future 5G networks is to maximize energy efficiency. Simultaneous wireless information and power transfer (SWIPT), which was initially proposed in \cite{4595260}, has rekindled the interest of researchers to explore more energy efficient networks. In \cite{4595260}, it was assumed that both  information and  energy could be extracted from the same radio frequency signals at the same time, which does not hold in practice. Motivated by this issue, two practical receiver architectures, namely time switching (TS) receiver and power splitting (PS) receiver, were proposed in a multi-input and multi-output (MIMO) system in~\cite{zhang2013mimo}. Since point-to-point communication systems with SWIPT are well established in the existing literature, recent research on SWIPT has focused on two common cooperative relaying systems: amplify-and-forward (AF) and decode-and-forward (DF). On the one hand, for AF relaying, a TS-based relaying protocol and a PS-based relaying protocol were proposed in~\cite{nasir2013relaying}. On the other hand, for DF relaying, a new antenna switching SWIPT protocol was proposed in \cite{krikidis2014low} to lower the implementation complexity. In \cite{ding2014wireless}, the application of SWIPT to DF cooperative networks with   randomly deployed  relays was investigated using stochastic geometry in a cooperative scenario with multiple source nodes and a single destination. A scenario in which multiple source-destination pairs are randomly deployed and communicate with each other via a single energy harvesting relay was considered in \cite{ding2013cooperative}.
\subsection{Motivation and Contributions}
One important advantage  of the NOMA concept is that it can squeeze a user with better channel conditions into a channel that is occupied by a user with worse channel conditions \cite{ding2014pairing}. For example, consider a downlink scenario in which there are two groups of users: 1) near users, which are close to the base station (BS) and have better channel conditions; and 2) far users, which are close to the edge of the cell controlled by the BS and therefore have worse channel conditions. While the spectral efficiency of NOMA is superior compared to orthogonal MA, the fact that the near users  co-exist with the far users   causes  performance degradation to the far users. In order to improve the reliability of the far users, an efficient method  was proposed in \cite{ding2014letter} by applying cooperative transmission to NOMA.  The key idea of this cooperative NOMA scheme is that the users that are close to the BS are used as relays to help the far  users with poor channel conditions. The advantage of implementing cooperative transmission in NOMA systems is that successive interference cancelation  is used at the near users and hence  the information of the far users is known by  these  near users. In this case, it is natural to consider the use of the near users as DF relays to transmit information to the far users.

In this paper, we consider this setting, but with the additional feature that the near users are energy constrained  and hence harvest energy from their received RF signals. To improve the reliability of the far NOMA users without draining the near users' batteries, we consider the application of SWIPT to NOMA, where SWIPT is performed at the near NOMA users. Therefore, the aforementioned two communication concepts, cooperative NOMA and SWIPT, can be naturally linked together, and a new spectrally  and energy efficient wireless multiple access protocol, namely, the cooperative SWIPT NOMA protocol,  is proposed in this paper. In order to investigate the impact of the locations of randomly deployed users on the performance of the proposed protocol, tools from stochastic geometry are used. Particularly, users are spatially randomly deployed in two groups via  homogeneous Poisson point processes (PPPs). Here, the near users are grouped together and randomly deployed in an area close to the BS. The far users are in the other group and are deployed close to the edge of the cell controlled by the BS.

Since NOMA is co-channel interference limited, it is important to combine  NOMA with conventional orthogonal MA technologies and realize a new hybrid MA network. For example, we can first group users in pairs to perform NOMA, and then use conventional time/frequency/code division MA to serve the different user pairs. Note that this hybrid MA scheme can effectively reduce the system complexity since fewer users are grouped together for the implementation of NOMA.
Based on the proposed protocol and the considered stochastic geometric  model, a natural question arises: which near NOMA user should help which far NOMA user?  To investigate the performance of one pair of selected NOMA users, three opportunistic user selection schemes are proposed, based on locations of users to  perform NOMA as follows: 1) random near user and random far user (RNRF) selection, where both the near and far users are randomly selected from the two groups; 2) nearest near user and nearest far user (NNNF) selection, where a near user and a far user closest to the BS are selected from the two groups; and 3) nearest near user and farthest far user (NNFF) selection, where a near user which is closest to the BS is selected and a far user which is farthest from the BS is selected. The insights obtained from these opportunistic user selection schemes provide guidance for the design of dynamic  user clustering algorithms, a topic beyond the scope of the paper.

The primary contributions of our paper are summarized as follows.

\begin{itemize}
  \item We propose a new SWIPT NOMA protocol to improve the reliability of the far users with the help of the near users without consuming extra energy. With this in mind, three user selection schemes are proposed by opportunistically taking into account the users' locations.
  \item We derive closed-form expressions for the outage probability at the near and far users, when considering the three proposed user selection schemes. In addition, we analyze the delay-sensitive throughput based on the outage probabilities of the near and far users.
  \item We derive the diversity gain of the three proposed selection schemes for the near and far users. We conclude that all three schemes have the same diversity order. For the far users, it is worth noting that the diversity gain of the proposed cooperative SWIPT NOMA  is the same as that of a conventional cooperative network without radio frequency energy harvesting.
  \item Comparing RNRF, NNNF, and NNFF, we confirm that NNNF achieves the lowest outage probability and the highest throughput for both the near and far users.
\end{itemize}

\subsection{Organization}
The rest of the paper is organized as follows. In Section \ref{System Model}, the network model for studying cooperative SWIPT NOMA is presented. In Section \ref{Analysis}, new analytical expressions are derived for the outage probability, diversity gain, and throughput when  the proposed selection schemes, RNRF, NNNF, and NNFF, are used. Numerical results are presented in Section \ref{Numerical Results}, which is followed by conclusion in Sections \ref{Conclusions}.
\section{Network Model}\label{System Model}
\begin{figure}[t!]
    \begin{center}
        \includegraphics[width=3.3 in]{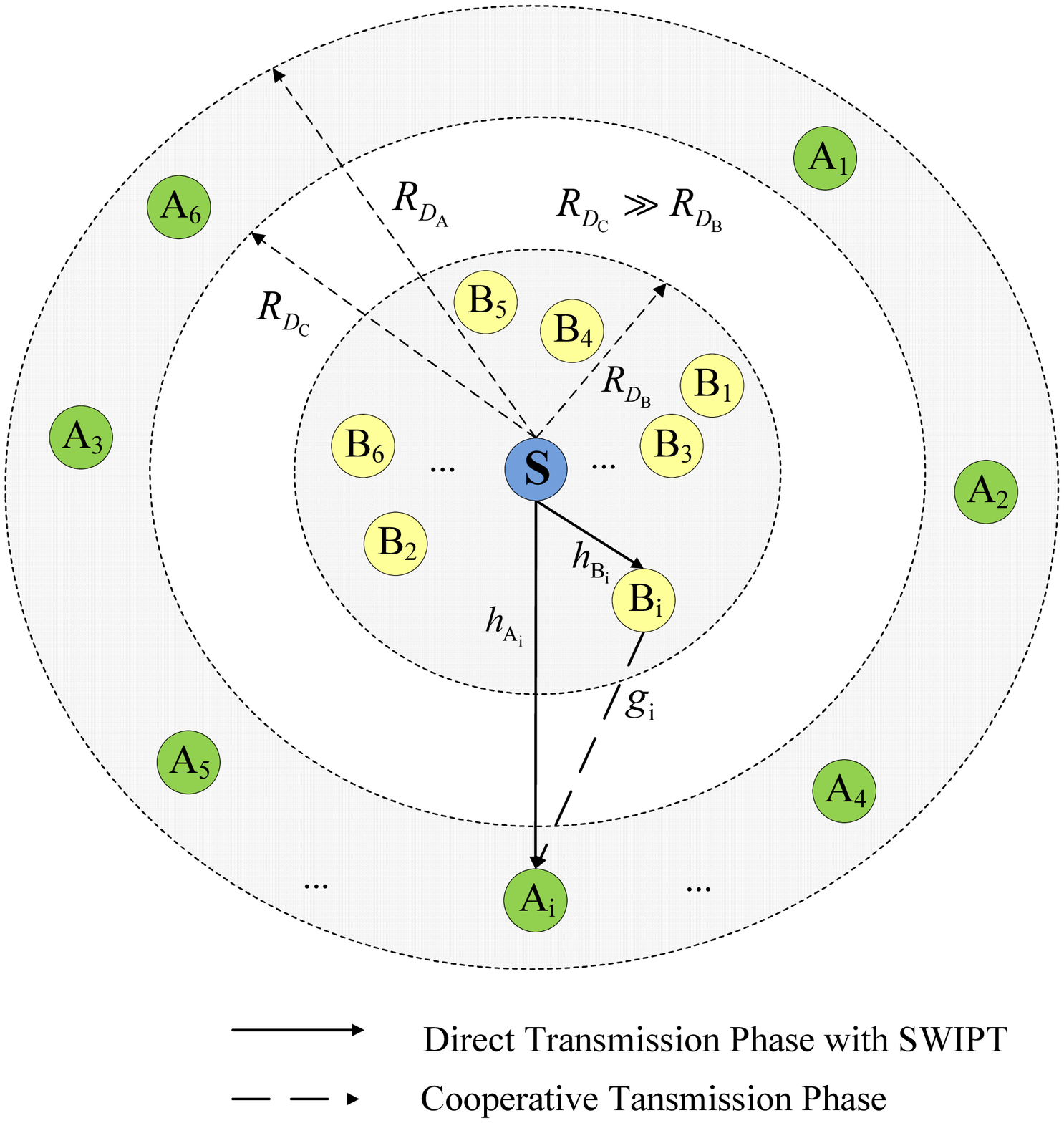}
        \caption{An illustration of a aownlink SWIPT NOMA system with a base station S (blue circle). The spatial distributions of the near users (yellow circles) and the far users (green circles) follow homogeneous PPPs.}
        \label{Fig1}
    \end{center}
\end{figure}

We consider a network with a single source $\mathrm{S}$ (i.e., the base station (BS)) and two groups of randomly deployed users $\{\mathrm{A_i}\}$ and $\{\mathrm{B_i}\}$. We assume that the users in  group $\{\mathrm{B_i}\}$ are deployed   within disc ${D_{\rm{B}}}$ with radius ${R_{{D_{\rm{B}}}}}$.  The far users $\{\mathrm{A_i}\}$ are deployed within   ring ${D_{\rm{A}}}$ with radius ${R_{{D_{\rm{C}}}}}$  and ${R_{{D_{\rm{A}}}}}$(assuming ${R_{{D_{\rm{C}}}}} \gg {R_{{D_{\rm{B}}}}}$), as shown in Fig. \ref{Fig1}. Note that the BS is located at the origin of both the disc ${D_{\rm{B}}}$ and the ring ${D_{\rm{A}}}$.   The locations of the near and far users are modeled as homogeneous PPPs $\Phi_{\kappa} $ ($\kappa\in \left\{ {{\rm{A,B}}} \right\}$) with densities $\lambda_{\Phi_\kappa }$. Here the near users are uniformly distributed within the disc and the far users are uniformly distributed within the ring. The number of users in ${R_{{D_{\kappa}}}}$, denoted by $N_\kappa$, follows a Poisson distribution $\Pr \left( {N_\kappa = k} \right) = (\mu _\kappa ^k/k!){e^{ - {\mu _\kappa }}}$, where ${\mu _\kappa }$ is the mean measure, i.e., ${\mu _{\rm{A}}} = \pi \left( {R_{{D_{\rm{A}}}}^2 - R_{{D_{\rm{C}}}}^2} \right){\lambda _{{\Phi _{\rm{A}}}}}$ and ${\mu _{\rm{B}}} = \pi R_{{D_{\rm{B}}}}^2{\lambda _{{\Phi _{\rm{B}}}}}$. All channels are assumed to be quasi-static Rayleigh fading, where the channel coefficients are constant for each transmission block but vary independently between different blocks. In the proposed network, we consider that the users in $\{\mathrm{B_i}\}$ are energy harvesting relays that harvest energy from the BS and forward the information to $\{\mathrm{A_i}\}$ using the harvested energy as their transmit powers. The DF strategy is applied at $\{\mathrm{B_i}\}$ and the cooperative NOMA system consists of two phases, detailed in the following. In this work, without loss of generality, it is assumed that the two phases have the same transmission periods, the same as in \cite{nasir2013relaying,ding2014wireless,ding2013cooperative}. It is worth pointing out that dynamic time allocation for the two phases may further improve the performance of the proposed cooperative NOMA scheme, but consideration of this issue is beyond the scope of the paper.

\subsection{Phase 1: Direct Transmission}
Prior to transmission, the two users denoted by $\mathrm{A_i}$ and $\mathrm{B_i}$, are selected to perform NOMA, where the selection criterion will be discussed in the next section. During the first phase, the BS sends two messages  ${p_{i1}}{x_{i1}} + {p_{i2}}{x_{i2}}$ to two selected users $\mathrm{A_i}$ and $\mathrm{B_i}$ based on NOMA~\cite{ding2014performance}, where $p_{i1}$ and $ p_{i2}$ are the power allocation coefficients and $x_{i1}$ and $ x_{i2}$ are the messages of $\mathrm{A_i}$ and $\mathrm{B_i}$, respectively. The observation at $\mathrm{A_i}$  is given by
\begin{align}\label{DT reveiving signal A}
{y_{\mathrm{A_i},1}} = \sqrt {{P_{\rm{S}}}}\sum\limits_{k \in \left\{ {1,2} \right\}} {{p_{ik}}{x_{ik}}} \frac{{{h_\mathrm{A_i}}}}{\sqrt  {1+d_{\rm{A_i}}^\alpha }}  + {n_{\mathrm{A_i},1}},
\end{align}
where ${P_{\rm{S}}}$ is the transmit power at the BS, $h_\mathrm{A_i}$ models the small-scale Rayleigh fading from the BS to $\mathrm{A_i}$ with ${h_\mathrm{A_i}} \sim \mathcal{CN}( {0,1})$, $n_{\mathrm{A_i},1}$ is additive Gaussian white noise (AWGN) at $\mathrm{A_i}$ with variance $\sigma _\mathrm{A_i}^2$, ${d_\mathrm{A_i}}$ is the distance between BS and $\mathrm{A_i}$, and $\alpha$ is the path loss exponent.

Without loss of generality, we assume that  ${\left| {{p_{i1}}} \right|^2} > {\left| {{p_{i2}}} \right|^2}$ with ${\left| {{p_{i1}}} \right|^2} + {\left| {{p_{i2}}} \right|^2} = 1$. The received signal to interference and noise ratio (SINR) at $\mathrm{A_i}$ to detect $x_{i1}$ is given by
\begin{align}\label{DT SNR A}
\gamma _{{\rm{S}},{{\rm{A}}_{\rm{i}}}}^{{x_{i1}}} = \frac{{\rho {{\left| {{h_{{{\rm{A}}_{\rm{i}}}}}} \right|}^2}{{\left| {{p_{i1}}} \right|}^2}}}{{\rho {{\left| {{p_{i2}}} \right|}^2}{{\left| {{h_{{{\rm{A}}_{\rm{i}}}}}} \right|}^2} +1+  d_{{{\rm{A}}_{\rm{i}}}}^\alpha }},
\end{align}
where $\rho  = \frac{{P_{\rm{S}}}}{{{\sigma ^2}}}$ is the transmit signal to noise radio (SNR) (assuming $\sigma _{\rm{A_i}}^2 = \sigma _{\rm{B_i}}^2 = {\sigma ^2}$).

We consider that the near users have rechargeable storage ability~\cite{nasir2013relaying} and power splitting \cite{zhang2013mimo} is applied to perform SWIPT. From the implementation point of view, this rechargeable storage unit can be a supercapacitor or a short-term high-efficiency battery \cite{krikidis2014low}. The power splitting approach is applied as  explained in the following: the observation at  $\mathrm{B_i}$ is divided into two parts. One part is used  for information decoding by directing the observation flow to the detection circuit and the remaining part is used for energy harvesting to powers $\mathrm{B_i}$ for helping  $\mathrm{A_i}$. Thus,
\begin{align}\label{DT reveiving signal B}
{y_{\mathrm{B_i},1}} = \sqrt {{P_{\rm{S}}}}\sum\limits_{k \in \left\{ {1,2} \right\}} {{p_{ik}}{x_{ik}}} \frac{{\sqrt {1 - \beta_i } {h_\mathrm{B_i}}}}{\sqrt {1+d_{\rm{B_i}}^\alpha }} + {n_{\mathrm{B_i},1}},
\end{align}
where $\beta_i$ is the power splitting coefficient which is detailed in  \eqref{theta_opp}, $h_\mathrm{B_i}$ models the small-scale Rayleigh fading from the BS to $\mathrm{B_i}$ with ${h_\mathrm{B_i}} \sim \mathcal{CN}( {0,1})$, $n_\mathrm{B_i}$ is AWGN at ${n_{\mathrm{B_i},1}}$ with variance $\sigma _\mathrm{B_i}^2$, and ${d_\mathrm{B_i}}$ is the distance between the BS and $\mathrm{B_i}$.  We use the bounded path loss model to ensure that the path loss is always larger than one even for small distances \cite{ding2014wireless}.

Applying NOMA, successive interference cancellation (SIC)\cite{cover2006elements} is carried out at $\mathrm{B_i}$. Particularly, $\mathrm{B_i}$ first decodes the message of $\mathrm{A_i}$, then subtracts this component from the received signal to detect its own information.  Therefore, the received SINR at $\mathrm{B_i}$ to detect $x_{i1}$ of $\mathrm{A_i}$ is given by
\begin{align}\label{DT SNR B x1}
\gamma _{{\rm{S}},{{\rm{B}}_{\rm{i}}}}^{{x_{i1}}} = \frac{{\rho {{\left| {{h_{{{\rm{B}}_{\rm{i}}}}}} \right|}^2}{{\left| {{p_{i1}}} \right|}^2}\left( {1 - {\beta _i}} \right)}}{{\rho {{\left| {{h_{{{\rm{B}}_{\rm{i}}}}}} \right|}^2}{{\left| {{p_{i2}}} \right|}^2}\left( {1 - {\beta _i}} \right) + 1 + d_{{{\rm{B}}_{\rm{i}}}}^\alpha }}.
\end{align}
The received SNR at $\mathrm{B_i}$ to detect $x_{i2}$ of $\mathrm{B_i}$ is given by
\begin{align}\label{DT SNR B x2}
\gamma _{{\rm{S}},{{\rm{B}}_{\rm{i}}}}^{{x_{i2}}} = \frac{{\rho {{\left| {{h_{{{\rm{B}}_{\rm{i}}}}}} \right|}^2}{{\left| {{p_{i2}}} \right|}^2}\left( {1 - {\beta _i}} \right)}}{{1 + d_{{{\rm{B}}_{\rm{i}}}}^\alpha }}.
\end{align}

The power splitting coefficient $\beta_i$ is used to determine the amount of harvested energy. Based on \eqref{DT SNR B x1}, the data rate supported by the channel from the BS to $\mathrm{B_i}$ for decoding $ x_{i1}$ is given by
\begin{align}\label{Rate x1}
{R_{{x_{i1}}}} = \frac{1}{2}\log \left( {1 + \frac{{\rho {{\left| {{h_{{{\rm{B}}_{\rm{i}}}}}} \right|}^2}{{\left| {{p_{i1}}} \right|}^2}\left( {1 - {\beta _i}} \right)}}{{\rho {{\left| {{h_{{{\rm{B}}_{\rm{i}}}}}} \right|}^2}{{\left| {{p_{i2}}} \right|}^2}\left( {1 - {\beta _i}} \right) + 1 + d_{{{\rm{B}}_{\rm{i}}}}^\alpha }}} \right).
\end{align}

We assume that the energy required to receive/process information is negligible compared to the energy required for information transmission~\cite{nasir2013relaying}. In this work, we apply the dynamic power splitting protocol which means that the power splitting coefficient ${\beta _i}$ is a variable and opportunistically tuned to support the relay transmission. Our aim is to first guarantee the detection of the message of the far NOMA user, $\mathrm{A_i}$, at the near NOMA user $\mathrm{B_i}$, then $\mathrm{B_i}$ can harvest the remaining energy. In this case, based on \eqref{Rate x1}, in order to ensure that $\mathrm{B_i}$ can successfully decode the information of $\mathrm{A_i}$, we have a rate, i.e., $R_1=R_{x_{i1}}$. Therefore,  the power splitting coefficient is set as follows:
\begin{align}\label{theta_opp}
{\beta _i} = \max \left\{ {0,1 - \frac{{{\tau _1}\left( {1 + d_{{{\rm{B}}_{\rm{i}}}}^\alpha } \right)}}{{\rho \left( {{{\left| {{p_{i1}}} \right|}^2} - {\tau _1}{{\left| {{p_{i2}}} \right|}^2}} \right){{\left| {{h_{{{\rm{B}}_{\rm{i}}}}}} \right|}^2}}}} \right\},
\end{align}
where ${\tau _1} = {2^{2{R_1}}} - 1$. Here $\beta _i=0$ means that all the energy is used for information decoding and no energy remains for energy harvesting.

Based on \eqref{DT reveiving signal B}, the energy harvested at $\mathrm{B_i}$ is given by
\begin{align}\label{EH B fixed}
{E_{\rm{B_i}}}=\frac{{T\eta {P_{\rm{S}}}\beta_i {{\left| {{h_{{{\rm{B}}_{\rm{i}}}}}} \right|}^2}}}{{2\left( {1 + d_{{{\rm{B}}_{\rm{i}}}}^\alpha } \right)}},
\end{align}
where $T$ is the time period for the entire transmission including the direct transmission phase and the cooperative transmission phase, and $\eta$ is the energy harvesting coefficient. We assume that the two phases have the same transmission period, and therefore, the transmit power at $\mathrm{B_i}$ can be expressed as follows:
\begin{align}\label{P B fixed}
{P_t} = \frac{{\eta {P_{\rm{S}}}\beta_i {{\left| {{h_{{{\rm{B}}_{\rm{i}}}}}} \right|}^2}}}{{1 + d_{{{\rm{B}}_{\rm{i}}}}^\alpha }}.
\end{align}

\subsection{Phase 2: Cooperative Transmission}
During this phase, $\mathrm{B_i}$ forwards $x_{i1}$ to $\mathrm{A_i}$ by using the harvested energy during the direct transmission phase. In this case, $\mathrm{A_i}$ observes
\begin{align}\label{CP reveiving signal A}
{y_{{{\rm{A}}_{\rm{i}}},2}} = \frac{{\sqrt {{P_t}} {x_{i1}}{g_{\rm{i}}}}}{{\sqrt {1 + d_{{{\rm{C}}_{\rm{i}}}}^\alpha } }} + {n_{{{\rm{A}}_{\rm{i}}},2}},
\end{align}
where $g_\mathrm{i}$ models the small-scale Rayleigh fading from $\mathrm{B_i}$ to $\mathrm{A_i}$ with $g_\mathrm{i} \sim \mathcal{CN}( {0,1})$, ${n_{\rm{A_i},2}}$ is AWGN at $\mathrm{A_i}$ with variance $\sigma _\mathrm{A_i}^2$, ${d_{{{\rm{C}}_{\rm{i}}}}} = \sqrt {d_{{{\rm{A}}_{\rm{i}}}}^2 + d_{{{\rm{B}}_{\rm{i}}}}^2 - 2{d_{{{\rm{A}}_{\rm{i}}}}}{d_{{{\rm{B}}_{\rm{i}}}}}\cos \left( {{\theta _i}} \right)} $ is the distance between $\mathrm{B_i}$ and $\mathrm{A_i}$, and ${\theta _i}$ denotes the angle $\angle {{\rm{A}}_{\rm{i}}}{\rm{S}}{{\rm{B}}_{\rm{i}}}$.

Based on \eqref{P B fixed} and \eqref{CP reveiving signal A}, the received SNR for $\mathrm{A_i}$ to detect $x_{i1}$ forwarded from $\mathrm{B_i}$ is given by
\begin{align}\label{CT SNR}
\gamma _{{{\rm{A}}_{\rm{i}}},{{\rm{B}}_{\rm{i}}}}^{{x_{i1}}} = \frac{{{P_t}{{\left| {{g_i}} \right|}^2}}}{{\left( {1 + d_{{{\rm{C}}_{\rm{i}}}}^\alpha } \right){\sigma ^2}}} = \frac{{\eta \rho {\beta_i} {{\left| {{h_{{{\rm{B}}_{\rm{i}}}}}} \right|}^2}{{\left| {{g_i}} \right|}^2}}}{{\left( {1 + d_{{{\rm{C}}_{\rm{i}}}}^\alpha } \right)\left( {1 + d_{{{\rm{B}}_{\rm{i}}}}^\alpha } \right)}}.
\end{align}

At the end of this phase, $\mathrm{A_i}$ combines the signals from the BS and $\mathrm{B_i}$ using maximal-ratio combining (MRC). Combining the SNR of the direct transmission phase \eqref{DT SNR A} and the SINR of the cooperative transmission phase \eqref{CT SNR}, we obtain the received SINR at $\mathrm{A_i}$ as follows:
\begin{align}\label{A SNR total}
\gamma _{{{\rm{A}}_{\rm{i}}},{\rm{MRC}}}^{{x_{i1}}} = \frac{{\rho {{\left| {{h_{{{\rm{A}}_{\rm{i}}}}}} \right|}^2}{{\left| {{p_{i1}}} \right|}^2}}}{{\rho {{\left| {{h_{{{\rm{A}}_{\rm{i}}}}}} \right|}^2}{{\left| {{p_{i2}}} \right|}^2} + 1 + d_{{{\rm{A}}_{\rm{i}}}}^\alpha }} + \frac{{\eta \rho {\beta _i}{{\left| {{h_{{{\rm{B}}_{\rm{i}}}}}} \right|}^2}{{\left| {{g_i}} \right|}^2}}}{{\left( {1 + d_{{{\rm{B}}_{\rm{i}}}}^\alpha } \right)\left( {1 + d_{{{\rm{C}}_{\rm{i}}}}^\alpha } \right)}}.
\end{align}

\section{Non-Orthogonal Multiple Access with User Selection}\label{Analysis}
In this section, the performance of three user selection schemes are characterized in the following.
\subsection{RNRF Selection Scheme}
In this scheme, the BS randomly selects a near user $\rm{B_i}$ and a far user $\rm{A_i}$. This selection scheme provides a fair opportunity for each user to access the source with the NOMA protocol. The advantage of this user selection scheme is that it does not require the knowledge of  instantaneous channel state information (CSI). To make meaningful conclusions, in the rest of the paper, we only focus on ${\beta _i} > 0$ and the number of near users and far users satisfy ${N_B} \ge 1,{N_A} \ge 1$.
\subsubsection{Outage Probability of the Near Users of RNRF}
In the NOMA protocol, an outage of $\mathrm{B_i}$ can occur for two reasons. The first is that $\mathrm{B_i}$ cannot detect $x_{i1}$. The second is that $\mathrm{B_i}$ can detect $x_{i1}$ but cannot detect $x_{i2}$. To guarantee that the NOMA protocol can be implemented, the condition $ {\left| {{p_{i1}}} \right|^2} - {\left| {{p_{i2}}} \right|^2}{\tau _1} > 0$ should be satisfied \cite{ding2014performance}. Based on this, the outage probability of $\mathrm{B_i}$ can be expressed as follows:
\begin{align}\label{OP B 1}
&{P_{{{\rm{B}}_{\rm{i}}}}} = \Pr \left( {\frac{{\rho {{\left| {{h_{{{\rm{B}}_{\rm{i}}}}}} \right|}^2}{{\left| {{p_{i1}}} \right|}^2}}}{{\rho {{\left| {{h_{{{\rm{B}}_{\rm{i}}}}}} \right|}^2}{{\left| {{p_{i2}}} \right|}^2} + 1 + d_{{{\rm{B}}_{\rm{i}}}}^\alpha }} < {\tau _1}} \right)\nonumber\\
& + \Pr \left( {\frac{{\rho {{\left| {{h_{{{\rm{B}}_{\rm{i}}}}}} \right|}^2}{{\left| {{p_{i1}}} \right|}^2}}}{{\rho {{\left| {{h_{{{\rm{B}}_{\rm{i}}}}}} \right|}^2}{{\left| {{p_{i2}}} \right|}^2} + 1 + d_{{{\rm{B}}_{\rm{i}}}}^\alpha }} > {\tau _1},\gamma _{{\rm{S}},{{\rm{B}}_{\rm{i}}}}^{{x_{i2}}} < {\tau _2}} \right),
\end{align}
where ${\tau _2} = {2^{2{R_2}}} - 1$ with $R_2$ being the target rate at which $\mathrm{B_i}$ can detect $x_{i2}$.

The following theorem provides the outage probability of the near users in RNRF for an arbitrary choice of $\alpha$.
\begin{theorem}\label{theorem:1}
\emph{Conditioned on the PPPs, the outage probability of the near users $\mathrm{B_i}$ can be approximated as follows:}
\begin{align}\label{OP B GC}
{P_{{{\rm{B}}_{\rm{i}}}}} \approx \frac{1}{2}\sum\limits_{n = 1}^N {{\omega _N}\sqrt {1 - {\phi _n}^2} \left( {1 - {e^{ - {c_n}{\varepsilon _{{{\rm{A}}_i}}}}}} \right)\left( {{\phi _n} + 1} \right)} ,
\end{align}
\emph{if} ${\varepsilon _{{{\rm{A}}_i}}} \ge {\varepsilon _{{{\rm{B}}_i}}}$, \emph{otherwise} ${P_{{{\rm{B}}_{\rm{i}}}}} =1$, \emph{where} $ {\varepsilon _{{{\rm{A}}_i}}} = \frac{{{\tau _1}}}{{\rho \left( {{{\left| {{p_{i1}}} \right|}^2} - {{\left| {{p_{i2}}} \right|}^2}{\tau _1}} \right)}}$ \emph{and} ${\varepsilon _{{{\rm{B}}_i}}} = \frac{{{\tau _2}}}{{\rho {{\left| {{p_{i2}}} \right|}^2}}}$, \emph{$N$ is a parameter to ensure a complexity-accuracy tradeoff}, ${c_n} = 1 + {\left( {\frac{{{R_{{D_{\rm{B}}}}}}}{2}\left( {{\phi _n} + 1} \right)} \right)^\alpha }$, ${\omega _N} = \frac{\pi }{N}$, \emph{and} ${\phi _n} = \cos \left( {\frac{{2n - 1}}{{2N}}\pi } \right)$.
\begin{proof}
Define ${X_i} = \frac{{{{\left| {{h_{{{\rm{A}}_{\rm{i}}}}}} \right|}^2}}}{{1 + d_{{{\rm{A}}_{\rm{i}}}}^\alpha }},{Y_i} = \frac{{{{\left| {{h_{{{\rm{B}}_{\rm{i}}}}}} \right|}^2}}}{{1 + d_{{{\rm{B}}_{\rm{i}}}}^\alpha }}$, and ${Z_i} = \frac{{{{\left| {{g_i}} \right|}^2}}}{{1 + d_{{{\rm{C}}_{\rm{i}}}}^\alpha }}$. Substituting \eqref{DT SNR B x1} and \eqref{DT SNR B x2} into \eqref{OP B 1},  the outage probability of the near users is given by
\begin{align}\label{OP B 1_1}
{P_{{{\rm{B}}_{\rm{i}}}}} = \Pr \left( {{Y_i} < {\varepsilon _{{{\rm{A}}_i}}}} \right) + \Pr \left( {{Y_i} > {\varepsilon _{{{\rm{A}}_i}}},{\varepsilon _{{{\rm{A}}_i}}} < {\varepsilon _{{{\rm{B}}_i}}}} \right).
\end{align}

If ${\varepsilon _{{{\rm{A}}_i}}} < {\varepsilon _{{{\rm{B}}_i}}}$,  the outage probability at the near users is always one.

For the case ${\varepsilon _{{{\rm{A}}_i}}} \ge {\varepsilon _{{{\rm{B}}_i}}}$, note that the users are deployed in ${D_{\rm{B}}}$ and ${D_{\rm{A}}}$ according to homogeneous PPPs. Therefore, the NOMA users are modeled as independently and identically distributed (i.i.d.) points in  ${D_{\rm{B}}}$ and ${D_{\rm{A}}}$, denoted by ${W_{{\kappa_i}}}$ $\left( {\kappa \in \left\{ {\mathrm{A},\mathrm{B}} \right\}} \right)$, which contain the location information about $\mathrm{A_i}$ and $\mathrm{B_i}$, respectively. The probability density functions (PDFs) of ${W_{{\mathrm{A_i}}}}$ and ${W_{{\mathrm{B_i}}}}$ are given by
\begin{align}\label{W_B}
{f_{{W_{{{\rm{B}}_{\rm{i}}}}}}}\left( {{\omega _{{{\rm{B}}_{\rm{i}}}}}} \right) = \frac{{{\lambda _{{\Phi _{\rm{B}}}}}}}{{{\mu _{{R_{{D_{\rm{B}}}}}}}}} = \frac{1}{{\pi R_{{D_{\rm{B}}}}^2}},
\end{align}
and
\begin{align}\label{W_A}
{f_{{W_{{{\rm{A}}_{\rm{i}}}}}}}\left( {{\omega _{{{\rm{A}}_{\rm{i}}}}}} \right) = \frac{{{\lambda _{{\Phi _{\rm{A}}}}}}}{{{\mu _{{R_{{D_{\rm{A}}}}}}}}} = \frac{1}{{\pi \left( {R_{{D_{\rm{A}}}}^2 - R_{{D_{\rm{C}}}}^2} \right)}},
\end{align}
respectively.

Therefore, for the case ${\varepsilon _{{{\rm{A}}_i}}} \ge {\varepsilon _{{{\rm{B}}_i}}}$, the cumulative distribution function (CDF) of ${Y_i}$ is given by
\begin{align}\label{Theta 2 CDF B}
{F_{Y_i}}\left( \varepsilon  \right) =& \int\limits_{D_{{\rm{B}}}} {\left( {1 - {e^{ - \left( {1 + d_{{{\rm{B}}_{\rm{i}}}}^\alpha } \right)\varepsilon }}} \right){f_{{W_{{{\rm{B}}_{\rm{i}}}}}}}\left( {{\omega _{{{\rm{B}}_{\rm{i}}}}}} \right)d{\omega _{{{\rm{B}}_{\rm{i}}}}}}\nonumber\\
=& \frac{2}{{R_{{D_{\rm{B}}}}^2}}\int_0^{{R_{{D_{\rm{B}}}}}} {\left( {1 - {e^{ - \left( {1 + {r^\alpha }} \right)\varepsilon }}} \right)rdr}.
\end{align}

For many communication scenarios $\alpha >2$, and it is challenging to obtain exact closed-from expressions for the above. In this case, we can use Gaussian-Chebyshev quadrature \cite{Hildebrand1987introduction} to find the approximation of \eqref{Theta 2 CDF B} as follows:
\begin{align}\label{GC quadrature CDF B}
{F_{Y_i}}\left( \varepsilon  \right) \approx \frac{1}{2}\sum\limits_{n = 1}^N {{\omega _N}\sqrt {1 - {\phi _n}^2} \left( {1 - {e^{ - {c_n}\varepsilon }}} \right)\left( {{\phi _n} + 1} \right)} .
\end{align}

Applying $ {\varepsilon _{{{\rm{A}}_i}}} \to \varepsilon$ into \eqref{GC quadrature CDF B}, \eqref{OP B GC} is obtained, and the proof of the theorem is completed.
\end{proof}
\end{theorem}

\begin{corollary}\label{corollary:1}
\emph{For the special case }$\alpha=2$, \emph{the outage probability of} $\mathrm{B_i}$ \emph{can be obtained as follows:}
\begin{align}\label{OP B alpha 2}
{\left. {{P_{{{\rm{B}}_{\rm{i}}}}}} \right|_{\alpha  = 2}} = 1 - \frac{{{e^{ - {\varepsilon _{{{\rm{A}}_i}}}}}}}{{R_{{D_{\rm{B}}}}^2{\varepsilon _{{{\rm{A}}_i}}}}} + \frac{{{e^{ - \left( {1 + R_{{D_{\rm{B}}}}^2} \right){\varepsilon _{{{\rm{A}}_i}}}}}}}{{R_{{D_{\rm{B}}}}^2{\varepsilon _{{{\rm{A}}_i}}}}},
\end{align}
\emph{if} ${\varepsilon _{{{\rm{A}}_i}}} \ge {\varepsilon _{{{\rm{B}}_i}}}$, \emph{otherwise} ${\left. {{P_{{{\rm{B}}_{\rm{i}}}}}} \right|_{\alpha  = 2}} =1$.
\begin{proof}
Based on \eqref{Theta 2 CDF B}, when $\alpha=2$ and after some manipulations, we can easily obtain
\begin{align}\label{CDF B alpha 2}
{\left. {{F_{Y_i}}\left( \varepsilon  \right)} \right|_{\alpha  = 2}} = 1 - \frac{{{e^{ - \varepsilon }}}}{{R_{{D_{\rm{B}}}}^2\varepsilon }} + \frac{{{e^{ - \left( {1 + R_{{D_{\rm{B}}}}^2} \right)\varepsilon }}}}{{R_{{D_{\rm{B}}}}^2\varepsilon }}.
\end{align}
Applying ${\varepsilon _{{{\rm{A}}_i}}} \to \varepsilon$ into \eqref{CDF B alpha 2}, \eqref{OP B alpha 2} can be obtained. The proof is completed.
\end{proof}
\end{corollary}

\subsubsection{Outage Probability of the Far Users of RNRF}
With the proposed cooperative SWIPT NOMA protocol, outage experienced by $\mathrm{A_{i}}$ can occur in two situations. The first is when $\mathrm{B_i}$ can detect $x_{i1}$ but the overall received SNR at $\mathrm{A_{i}}$ cannot support the targeted rate. The second is when  neither $\mathrm{A_{i}}$ nor $\mathrm{B_i}$ can detect $x_{i1}$. Based on this, the outage probability can be expressed as follows:
\begin{align}\label{OP A 1_1}
{P_{{{\rm{A}}_{\rm{i}}}}} =& \Pr \left( {\gamma _{{{\rm{A}}_{\rm{i}}},{\rm{MRC}}}^{{x_{i1}}} < {\tau _1},{{\left. {\gamma _{{\rm{S}},{{\rm{B}}_{\rm{i}}}}^{{x_{i1}}}} \right|}_{{\beta _i} = 0}} > {\tau _1}} \right)\nonumber\\
& + \Pr \left( {\gamma _{{\rm{S}},{{\rm{A}}_{\rm{i}}}}^{{x_{i1}}} < {\tau _1},{{\left. {\gamma _{{\rm{S}},{{\rm{B}}_{\rm{i}}}}^{{x_{i1}}}} \right|}_{{\beta _i} = 0}} < {\tau _1}} \right).
\end{align}

The following theorem provides the outage probability of the far users in RNRF for an arbitrary choice of $\alpha$.
\begin{theorem}\label{theorem:2}
\emph{Conditioned on the PPPs, and assuming ${R_{{D_{\rm{C}}}}} \gg {R_{{D_{\rm{B}}}}}$, the outage probability of} $\mathrm{A_i}$ \emph{can be approximated as follows:}
\begin{align}\label{OP A GC}
&{P_{{{\rm{A_i}}}}} \approx   \zeta_1 \sum\limits_{n = 1}^N {\left( {{\phi _n} + 1} \right)\sqrt {1 - {\phi _n}^2} } {c_n}\sum\limits_{k = 1}^K {\sqrt {1 - \psi _k^2} {s_k}{{\left( {1 + s_k^\alpha } \right)}^2}}\nonumber\\
&\times\sum\limits_{m = 1}^M {\sqrt {1 - \varphi _m^2} {e^{ - \left( {1 + s_k^\alpha } \right){t_m}}}{\chi _{{t_m}}}\left( {\ln \frac{{{\chi _{{t_m}}}\left( {1 + s_k^\alpha } \right)}}{{\eta \rho }}{c_n} + 2{c_0}} \right)}\nonumber\\
&+ {a_1}\sum\limits_{n = 1}^N {\sqrt {1 - {\phi _n}^2} {c_n}\left( {{\phi _n} + 1} \right)} \sum\limits_{k = 1}^K {\sqrt {1 - {\psi _k}^2} (1 + s_k^\alpha ){s_k}},
\end{align}
\emph{where $M$ and $K$ are parameters to ensure a complexity-accuracy tradeoff}, ${\zeta _1} =  - \frac{{{\varepsilon _{{{\rm{A}}_i}}}{R_{{D_{{{\rm{B}}_{\rm{i}}}}}}}{\omega _N}{\omega _K}{\omega _M}}}{{8\left( {{R_{{D_{{{\rm{A}}_{\rm{i}}}}}}} + {R_{{D_{{{\rm{C}}_{\rm{i}}}}}}}} \right)\eta \rho }},{\chi _{{t_m}}} = {\tau _1} - \frac{{\rho {t_m}{{\left| {{p_{i1}}} \right|}^2}}}{{\rho {t_m}{{\left| {{p_{i2}}} \right|}^2} + 1}},{t_m} = \frac{{{\varepsilon _{{{\rm{A}}_i}}}}}{2}\left( {{\varphi _m} + 1} \right),{\omega _M} = \frac{\pi }{M},{\varphi _m} = \cos \left( {\frac{{2m - 1}}{{2M}}\pi } \right), {s_k} = \frac{{{R_{{D_{\rm{A}}}}} - {R_{{D_{\rm{C}}}}}}}{2}\left( {{\psi _k} + 1} \right) + {R_{{D_{\rm{C}}}}},{\omega _K} = \frac{\pi }{K},{\psi _k} = \cos \left( {\frac{{2k - 1}}{{2K}}\pi } \right)$, ${c_0} =  - \frac{{\varphi \left( 1 \right)}}{2} - \frac{{\varphi \left( 2 \right)}}{2}$, \emph{and} ${a_1} = \frac{{{\omega _K}{\omega _N}\varepsilon _{{{\rm{A}}_1}}^2}}{{2\left( {{R_{{D_{\rm{A}}}}} + {R_{{D_{\rm{C}}}}}} \right)}}$.
\begin{proof}
See Appendix~A.
\end{proof}
\end{theorem}

\begin{corollary}\label{corollary:2}
\emph{For the special case }$\alpha=2$, \emph{the outage probability of} $\mathrm{A_i}$ \emph{can be simplified as follows:}
\begin{align}\label{OP A GC alpha 2}
{\left. {{P_{{{\rm{A}}_{\rm{i}}}}}} \right|_{\alpha  = 2}}& \approx  {\zeta _2}\sum\limits_{k = 1}^K {\sqrt {1 - \psi _k^2} {s_k}{{\left( {1 + s_k^2} \right)}^2}} \sum\limits_{m = 1}^M {\sqrt {1 - \varphi _m^2} }  \nonumber\\
&\times{\chi _{{t_m}}}{e^{ - \left( {1 + s_k^2} \right){t_m}}}\left( {\ln \frac{{{\chi _{{t_m}}}\left( {1 + s_k^2} \right)}}{{\eta \rho }}{c_n} + {b_0}} \right) \nonumber\\
& + \left( {1 - \frac{{{e^{ - \left( {1 + R_{{D_{\rm{C}}}}^2} \right){\varepsilon _{{{\rm{A}}_i}}}}}}}{{{\varepsilon _{{{\rm{A}}_i}}}\left( {R_{{D_{\rm{A}}}}^2 - R_{{D_{\rm{C}}}}^2} \right)}} + \frac{{{e^{ - \left( {1 + R_{{D_{\rm{A}}}}^2} \right){\varepsilon _{{{\rm{A}}_i}}}}}}}{{{\varepsilon _{{{\rm{A}}_i}}}\left( {R_{{D_{\rm{A}}}}^2 - R_{{D_{\rm{C}}}}^2} \right)}}} \right)\nonumber\\
&\times\left( {1 - \frac{{{e^{ - {\varepsilon _{{{\rm{A}}_i}}}}}}}{{R_{{D_{\rm{B}}}}^2{\varepsilon _{{{\rm{A}}_i}}}}} + \frac{{{e^{ - \left( {1 + R_{{D_{\rm{B}}}}^2} \right){\varepsilon _{{{\rm{A}}_i}}}}}}}{{R_{{D_{\rm{B}}}}^2{\varepsilon _{{{\rm{A}}_i}}}}}} \right),
\end{align}
\emph{where} ${\zeta _2}= -\frac{{{\omega _K}{\omega _M}{\varepsilon _{{{\rm{A}}_i}}}\left( {R_{{D_{\rm{B}}}}^2 + 2} \right)}}{{8\left( {{R_{{D_{\rm{A}}}}} + {R_{{D_{\rm{C}}}}}} \right)\eta \rho }}$ and ${b_0} = \frac{{{{\left( {1 + R_{{D_{\rm{B}}}}^2} \right)}^2}\ln \left( {1 + R_{{D_{\rm{B}}}}^2} \right)}}{{2R_{{D_{\rm{B}}}}^2}} + \left( {R_{{D_{\rm{B}}}}^2 + 2} \right)\left( {{c_0} - \frac{1}{4}} \right)$.
\begin{proof}
See Appendix~B.
\end{proof}
\end{corollary}
\subsubsection{Diversity Analysis of RNRF}
To obtain further insights into the derived outage probability, we provide a diversity analysis of both the near and far users of RNRF.

\emph{Near users:} For the near users, based on the analytical results, we carry out high SNR approximations as follows. When $\varepsilon  \to 0$, a high SNR approximation of \eqref{GC quadrature CDF B} with $1 - {e^{ - x}} \approx x$ is given by
\begin{align}\label{GC quadrature CDF B appox}
{F_{{Y_i}}}\left( \varepsilon  \right) \approx \frac{1}{2}\sum\limits_{n = 1}^N {{\omega _N}\sqrt {1 - {\phi _n}^2} {c_n}{\varepsilon _{{{\rm{A}}_i}}}\left( {{\phi _n} + 1} \right)}.
\end{align}

The diversity gain is defined as follows:
\begin{align}\label{Diversity random}
d =  - \mathop {\lim }\limits_{\rho  \to \infty } \frac{{{\log {P}}\left( \rho  \right)}}{{\log \rho }}.
\end{align}

Substituting \eqref{GC quadrature CDF B appox} into \eqref{Diversity random}, we obtain that the diversity gain for the near users is one, which means that using NOMA with energy harvesting will not decrease the diversity gain.

\emph{Far users:} For the far users, substituting \eqref{OP A GC} into \eqref{Diversity random}, we obtain
\begin{align}\label{Diversity random1}
d =& - \mathop {\lim }\limits_{\rho  \to \infty } \frac{{\log \left( { - \frac{1}{{{\rho ^2}}}\log \frac{1}{\rho }} \right)}}{{\log \rho }}\nonumber\\
=&- \mathop {\lim }\limits_{\rho  \to \infty } \frac{{\log \log \rho  - \log {\rho ^2}}}{{\log \rho }}=2 .
\end{align}

As we can see from \eqref{Diversity random1}, the diversity gain of RNRF is two, which is the same as that of the conventional cooperative network \cite{laneman2004cooperative}. This result indicates that using NOMA with an energy harvesting relay will not affect the diversity gain. In addition, we see that at high SNRs, the dominant factor for the outage probability is ${\frac{1}{{{\rho ^2}}}\ln \rho }$. Therefore we conclude that the outage probability of using NOMA with SWIPT decays at a rate of $\frac{{\ln SNR}}{{SN{R^2}}}$. However, for a conventional cooperative system without energy harvesting, a faster decreasing rate of $\frac{1}{{SN{R^2}}}$ can be achieved.

\subsubsection{System Throughput in Delay-Sensitive Transmission Mode of RNRF}
In this paper, we will focus on the delay-sensitive throughput. In this mode, the transmitter sends information at a fixed rate and the throughput is determined by evaluating the outage probability.

Based on the analytical results for the outage probability of the near and far users, the system throughput of RNRF in the delay-sensitive transmission mode is given by
\begin{align}\label{throughput random}
{R_{{\tau _{{\rm{RNRF}}}}}} = \left( {1 - {P_{{{\rm{A}}_{\rm{i}}}}}} \right){R_1} + \left( {1 - {P_{{{\rm{B}}_{\rm{i}}}}}} \right){R_2},
\end{align}
where ${P_{{{\rm{A}}_{\rm{i}}}}}$ and ${P_{{{\rm{B}}_{\rm{i}}}}}$ are obtained from \eqref{OP A GC} and \eqref{OP B GC}, respectively.
\subsection{NNNF Selection Scheme}
In this subsection, we characterize the performance of NNNF, which exploits the users' CSI opportunistically. We first select a user within the disc ${D_{\rm{B}}}$ which has the shortest distance to the BS as the near NOMA user (denoted by $\rm{B_{i^*}}$). This is because the near users also act as energy harvesting relays to help  the far users. The NNNF scheme can enable the selected near user to harvest more energy. Then we select a user within the ring ${D_{\rm{A}}}$ which has the shortest distance to the BS as the far NOMA user (denoted by $\rm{A_{i^*}}$). The advantage of the NNNF scheme is that it can minimize the outage probability of both the near and far users.
\subsubsection{Outage Probability of the Near Users of NNNF}
Using the same definition of the outage probability as the near users of NOMA, we can characterize the outage probability of the near users of NNNF.

The following theorem provides the outage probability of the near users of NNNF for an arbitrary choice of $\alpha$.
\begin{theorem}\label{theorem:3}
\emph{Conditioned on the PPPs, the outage probability of ${{{\rm{B}}_{\rm{i^*}}}}$ can be approximated  as follows:}
\begin{align}\label{OP Nearest B GC}
{P_{{{\rm{B}}_{{{\rm{i}}^{\rm{*}}}}}}} \approx {b_1}\sum\limits_{n = 1}^N {\sqrt {1 - {\phi _n}^2} \left( {1 - {e^{ - \left( {1 + c_{n*}^\alpha } \right){\varepsilon _{{{\rm{A}}_i}}}}}} \right){c_{n*}}{e^{ - \pi {\lambda _{{\Phi _{\rm{B}}}}}c_{n*}^2}}},
\end{align}
\emph{if} ${\varepsilon _{{{\rm{A}}_i}}} \ge {\varepsilon _{{{\rm{B}}_i}}}$, \emph{otherwise} ${P_{{{\rm{B}}_{{{\rm{i}}^{\rm{*}}}}}}} =1$, \emph{where} ${c_{n*}} = \frac{{{R_{{D_{\rm{B}}}}}}}{2}\left( {{\phi _n} + 1} \right)$, ${b_1} = \frac{{{\xi _{\rm{B}}}{\omega _N}{R_{{D_{\rm{B}}}}}}}{2}$, \emph{and} ${\xi _{\rm{B}}} = \frac{{2\pi {\lambda _{{\Phi _{\rm{B}}}}}}}{{1 - {e^{ - \pi {\lambda _{{\Phi _{\rm{B}}}}}R_{{D_{\rm{B}}}}^2}}}}$.
\begin{proof}
Similar to \eqref{OP B 1_1}, the outage probability of ${{\rm{B}}_{{\rm{i^*}}}}$ can be expressed as follows:
\begin{align}\label{OP Nearest B 1}
{P_{{{\rm{B}}_{{{\rm{i}}^{\rm{*}}}}}}} = \Pr \left( {\left. {{Y_{{i^*}}} < {\varepsilon _{{{\rm{A}}_i}}}} \right|{N_{\rm{B}}} \ge 1} \right) = {F_{Y_{i^*}}}\left( {{\varepsilon _{{{\rm{A}}_i}}}} \right),
\end{align}
where  ${Y_{{i^*}}} = \frac{{{{\left| {{h_{{{\rm{B}}_{\rm{i}}}}}} \right|}^2}}}{{1 + d_{{{\rm{B}}_{{\rm{i^*}}}}}^\alpha }}$ and $ d_{{{\rm{B}}_{\rm{i^*}}}}$ is the distance from the nearest ${{{\rm{B}}_{\rm{i^*}}}}$ to the BS.

The CDF of ${Y_{{i^*}}}$ can be written as follows:
\begin{align}\label{CDF Nearest B}
{F_{Y_{i^*}}}\left( \varepsilon  \right) = \int_0^{{R_{{D_{\rm{B}}}}}} {\left( {1 - {e^{ - \left( {1 + r_{\rm{B}}^\alpha } \right)\varepsilon }}} \right)} {f_{{d_{{{\rm{B}}_{\rm{i}}}*}}}}\left( {{r_{\rm{B}}}} \right)d{r_{\rm{B}}},
\end{align}
where ${f_{{d_{{{\rm{B}}_{\rm{i}}}*}}}}$ is the PDF of the shortest distance from ${{{\rm{B}}_{\rm{i^*}}}}$ to the BS.

The probability $\Pr \left\{ {\left. {{d_{{{\rm{B}}_{\rm{i}}}*}} > r} \right|{N_{\rm{B}}} \ge 1} \right\}$ conditioned on ${{N_{\rm{B}}} \ge 1}$ is  the event that there is no point located in the disc. Therefore we can express this probability as follows:
\begin{align}\label{CDF Nearest B 1}
&\Pr \left\{ {\left. {{d_{{{\rm{B}}_{\rm{i}}}*}} > r} \right|{N_{\rm{B}}} \ge 1} \right\}\nonumber\\
=& \frac{{\Pr \left\{ {{d_{{{\rm{B}}_{\rm{i}}}*}} > r} \right\} - \Pr \left\{ {{d_{{{\rm{B}}_{\rm{i}}}*}} > r,{N_{\rm{B}}} = 0} \right\}}}{{\Pr \left\{ {{N_{\rm{B}}} \ge 1} \right\}}}\nonumber\\
=& \frac{{{e^{ - \pi {\lambda _{{\Phi _{\rm{B}}}}}{r^2}}} - {e^{ - \pi {\lambda _{{\Phi _{\rm{B}}}}}R_{{D_{\rm{B}}}}^2}}}}{{1 - {e^{ - \pi {\lambda _{{\Phi _{\rm{B}}}}}R_{{D_{\rm{B}}}}^2}}}}.
\end{align}
Then the corresponding PDF of ${{{\rm{B}}_{\rm{i^*}}}}$ is given by
\begin{align}\label{PDF Nearest B}
{f_{{d_{{{\rm{B}}_{\rm{i}}}*}}}}\left( {{r_{\rm{B}}}} \right) = {\xi _{\rm{B}}}{r_{\rm{B}}}{e^{ - \pi {\lambda _{{\Phi _{\rm{B}}}}}r_{\rm{B}}^2}}.
\end{align}
Substituting \eqref{PDF Nearest B} into \eqref{CDF Nearest B}, we obtain
\begin{align}\label{CDF Nearest B 2}
{F_{Y_{i^*}}}\left( \varepsilon  \right) = {\xi _{\rm{B}}}\int_0^{{R_{{D_{\rm{B}}}}}} {\left( {1 - {e^{ - \left( {1 + r_{\rm{B}}^\alpha } \right)\varepsilon }}} \right)} {r_{\rm{B}}}{e^{ - \pi {\lambda _{{\Phi _{\rm{B}}}}}r_{\rm{B}}^2}}d{r_{\rm{B}}}.
\end{align}
Applying the Gaussian-Chebyshev quadrature approximation to \eqref{GC quadrature CDF B}, we obtain
\begin{align}\label{CDF Nearest B GC}
&{F_{Y_{i^*}}}\left( \varepsilon  \right) \approx \frac{{{\xi _{\rm{B}}}{\omega _N}{R_{{D_{\rm{B}}}}}}}{2}\nonumber\\
&\times\sum\limits_{n = 1}^N {\sqrt {1 - {\phi _n}^2} \left( {1 - {e^{ - \left( {1 + c_{n*}^\alpha } \right)\varepsilon }}} \right){c_{n*}}{e^{ - \pi {\lambda _{{\Phi _{\rm{B}}}}}c_{n*}^2}}}.
\end{align}
Applying ${\varepsilon _{{{\rm{A}}_i}}} \to \varepsilon$, we obtain the approximate outage probability of ${{{\rm{B}}_{\rm{i^*}}}}$ in \eqref{OP Nearest B GC}.
\end{proof}
\end{theorem}

Based on \eqref{CDF Nearest B 2} and after some manipulations, the following corollary can be obtained.
\begin{corollary}\label{corollary:3}
\emph{For the special case} $\alpha=2$,  \emph{the outage probability of $\rm{B_{i^*}}$ can be expressed as follows:}
\begin{align}\label{OP Nearest B GC alpha 2}
{\left. {{P_{{{\rm{B}}_{{{\rm{i}}^{\rm{*}}}}}}}} \right|_{\alpha  = 2}} =& \frac{{{\xi _{\rm{B}}}\left( {{e^{ - R_{{D_{\rm{B}}}}^2\left( {\pi {\lambda _{{\Phi _{\rm{B}}}}} + {\varepsilon _{{{\rm{A}}_i}}}} \right) - {\varepsilon _{{{\rm{A}}_i}}}}} - {e^{ - {\varepsilon _{{{\rm{A}}_i}}}}}} \right)}}{{2\left( {\pi {\lambda _{{\Phi _{\rm{B}}}}} + {\varepsilon _{{{\rm{A}}_i}}}} \right)}}\nonumber\\
&- \frac{{{\xi _{\rm{B}}}\left( {{e^{ - \pi {\lambda _{{\Phi _{\rm{B}}}}}R_{{D_{\rm{B}}}}^2}} - 1} \right)}}{{2\pi {\lambda _{{\Phi _{\rm{B}}}}}}},
\end{align}
\end{corollary}
\emph{if} ${\varepsilon _{{{\rm{A}}_i}}} \ge {\varepsilon _{{{\rm{B}}_i}}}$, \emph{otherwise} ${\left. {{P_{{{\rm{B}}_{{{\rm{i}}^{\rm{*}}}}}}}} \right|_{\alpha  = 2}}  =1$.

\subsubsection{Outage Probability of the Far Users of NNNF}
Using the same definition of the outage probability for the far users of NOMA, and similar to \eqref{OP A 1_1}, we can characterize the outage probability of the far users in NNNF.
The following theorem provides the outage probability of the far users in NNNF for an arbitrary choice of $\alpha$.
\begin{theorem}\label{theorem:4}
\emph{Conditioned on the PPPs and assuming ${R_{{D_{\rm{C}}}}} \gg {R_{{D_{\rm{B}}}}}$, the outage probability of} $\mathrm{A_{i^*}}$ \emph{can be approximated  as follows:}
\begin{align}\label{OP Nearest A GC}
&{P_{{{\rm{A}}_{{{\rm{i}}^{\rm{*}}}}}}} \approx {\varsigma ^*}\sum\limits_{n = 1}^N {\sqrt {1 - {\phi _n}^2} \left( {1 + c_{n*}^\alpha } \right){c_{n*}}{e^{ - \pi {\lambda _{{\Phi _{\rm{B}}}}}c_{n*}^2}}}\nonumber\\
&\times \sum\limits_{k = 1}^K {\sqrt {1 - \psi _k^2} {{\left( {1 + s_k^\alpha } \right)}^2}{s_k}{e^{ - \pi {\lambda _{{\Phi _{\rm{A}}}}}\left( {s_k^2 - R_{{D_{\rm{C}}}}^2} \right)}}} \sum\limits_{m = 1}^M {\sqrt {1 - \varphi _m^2} }\nonumber\\
&\times{e^{ - \left( {1 + s_k^\alpha } \right){t_m}}}{\chi _{{t_m}}}\left( {\ln \frac{{{\chi _{{t_m}}}\left( {1 + s_k^\alpha } \right)\left( {1 + c_{n*}^\alpha } \right)}}{{\eta \rho }} + 2{c_0}} \right)\nonumber\\
& + {b_2}{b_3}\sum\limits_{k = 1}^K {\sqrt {1 - {\psi _k}^2} (1 + s_k^\alpha ){s_k}{e^{ - \pi {\lambda _{{\Phi _{\rm{A}}}}}s_k^2}}} \nonumber\\
&\times \sum\limits_{n = 1}^N {\left( {\sqrt {1 - {\phi _n}^2} \left( {1 + c_{n*}^\alpha } \right){c_{n*}}{e^{ - \pi {\lambda _{{\Phi _{\rm{B}}}}}c_{n*}^2}}} \right)},
\end{align}
\emph{where} ${\varsigma ^*} =  - \frac{{{\xi _{\rm{B}}}{\xi _{\rm{A}}}{\omega _N}{\omega _K}{\omega _M}{\varepsilon _{{{\rm{A}}_i}}}{R_{{D_{\rm{B}}}}}\left( {{R_{{D_{\rm{A}}}}} - {R_{{D_{\rm{C}}}}}} \right)}}{{8\eta \rho }}$, ${b_2} = \frac{{{\xi _{\rm{A}}}{e^{\pi {\lambda _{{\Phi _{\rm{A}}}}}R_{{D_{\rm{C}}}}^2}}{\omega _K}{\varepsilon _{{{\rm{A}}_i}}}}}{{{R_{{D_{\rm{A}}}}} + {R_{{D_{\rm{C}}}}}}}$, \emph{and} ${b_3} = \frac{{{\xi _{\rm{B}}}{\omega _N}{R_{{D_{\rm{B}}}}}{\varepsilon _{{{\rm{A}}_i}}}}}{2}$.
\begin{proof}
See Appendix~C.
\end{proof}
\end{theorem}
\begin{corollary}\label{corollary:4}
\emph{For the special case} $\alpha=2$, \emph{the outage probability of} $\mathrm{A_{i^*}}$ \emph{can be simplified as} \eqref{OP Nearest A GC alpha 2} \emph{at the top of the following page}.
\begin{figure*}[!t]
\normalsize
\begin{align}\label{OP Nearest A GC alpha 2}
&{\left. {P_{{{\rm{A}}_{{\rm{i^*}}}}}} \right|_{\alpha  = 2}} \approx  {\varsigma ^*}\sum\limits_{n = 1}^N {\sqrt {1 - {\phi _n}^2} \left( {1 + c_{n*}^2} \right){c_{n*}}{e^{ - \pi {\lambda _{{\Phi _{\rm{B}}}}}c_{n*}^2}}} \sum\limits_{k = 1}^K {\sqrt {1 - \psi _k^2} {{\left( {1 + s_k^2} \right)}^2}{s_k}{e^{ - \pi {\lambda _{{\Phi _{\rm{A}}}}}\left( {s_k^2 - R_{{D_{\rm{C}}}}^2} \right)}}} \nonumber\\
&\times \sum\limits_{m = 1}^M {\sqrt {1 - \varphi _m^2} \left( {{e^{ - \left( {1 + s_k^2} \right){t_m}}}{\chi _{{t_m}}}\left( {\ln \frac{{{\chi _{{t_m}}}\left( {1 + s_k^2} \right)\left( {1 + c_{n*}^2} \right)}}{{\eta \rho }} + 2{c_0}} \right)} \right)}  \nonumber\\
&+ \frac{{{\xi _{\rm{A}}}{e^{\pi {\lambda _{{\Phi _{\rm{A}}}}}R_{{D_{\rm{C}}}}^2}}}}{2}\left( {\frac{{{e^{ - {\varepsilon _{{{\rm{A}}_i}}}}}}}{{\pi {\lambda _{{\Phi _{\rm{A}}}}} + {\varepsilon _{{{\rm{A}}_i}}}}}\left( {{e^{ - R_{{D_{\rm{A}}}}^2\left( {\pi {\lambda _{{\Phi _{\rm{A}}}}} + {\varepsilon _{{{\rm{A}}_i}}}} \right)}} - {e^{ - R_{{D_{\rm{C}}}}^2\left( {\pi {\lambda _{{\Phi _{\rm{A}}}}} + {\varepsilon _{{{\rm{A}}_i}}}} \right)}}} \right) - \frac{{\left( {{e^{ - \pi {\lambda _{{\Phi _{\rm{A}}}}}R_{{D_{\rm{A}}}}^2}} - {e^{ - \pi {\lambda _{{\Phi _{\rm{A}}}}}R_{{D_{\rm{C}}}}^2}}} \right)}}{{\pi {\lambda _{{\Phi _{\rm{A}}}}}}}} \right)\nonumber\\
&\times \frac{{{\xi _{\rm{B}}}}}{2}\left( {\frac{{{e^{ - R_{{D_{\rm{B}}}}^2\left( {\pi {\lambda _{{\Phi _{\rm{B}}}}} + {\varepsilon _{{{\rm{A}}_i}}}} \right) - {\varepsilon _{{{\rm{A}}_i}}}}} - {e^{ - {\varepsilon _{{{\rm{A}}_i}}}}}}}{{\pi {\lambda _{{\Phi _{\rm{B}}}}} + {\varepsilon _{{{\rm{A}}_i}}}}} - \frac{{{e^{ - \pi {\lambda _{{\Phi _{\rm{B}}}}}R_{{D_{\rm{B}}}}^2}} - 1}}{{\pi {\lambda _{{\Phi _{\rm{B}}}}}}}} \right).
\end{align}
\hrulefill \vspace*{0pt}
\end{figure*}
\begin{proof}
For the special case $\alpha=2$, after some manipulations, we can express \eqref{CDF Nearest A 1} as follows:
\begin{align}\label{CDF Nearest A alpha 2}
&{\left. {{F_{X_{i^*}}}\left( \varepsilon  \right)} \right|_{\alpha  = 2}} =  - \frac{{{\xi _{\rm{A}}}\left( {{e^{\pi {\lambda _{{\Phi _{\rm{A}}}}}\left( {R_{{D_{\rm{C}}}}^2 - R_{{D_{\rm{A}}}}^2} \right)}} - 1} \right)}}{{2\pi {\lambda _{{\Phi _{\rm{A}}}}}}}\nonumber\\
& + \frac{{{\xi _{\rm{A}}}{e^{\pi {\lambda _{{\Phi _{\rm{A}}}}}R_{{D_{\rm{C}}}}^2}}{e^{ - \varepsilon }}}}{{2\left( {\pi {\lambda _{{\Phi _{\rm{A}}}}} + \varepsilon } \right)}}\left( {{e^{ - R_{{D_{\rm{A}}}}^2\left( {\pi {\lambda _{{\Phi _{\rm{A}}}}} + \varepsilon } \right)}} - {e^{ - R_{{D_{\rm{C}}}}^2\left( {\pi {\lambda _{{\Phi _{\rm{A}}}}} + \varepsilon } \right)}}} \right).
\end{align}
Based on \eqref{Theta 2 Nearest A 1}, combining \eqref{CDF Nearest A alpha 2} and \eqref{OP Nearest B GC alpha 2}, and setting $\alpha=2$ into \eqref{theta1 nearest 3}, we can obtain \eqref{OP Nearest A GC alpha 2}. The proof is completed.
\end{proof}
\end{corollary}

\subsubsection{Diversity Analysis of NNNF}
Similarly, we provide diversity analysis of both the near and far users of NNNF.

\emph{Near users:}
For the near users, based on the analytical results, we carry out the high SNR approximation as follows. When $\varepsilon  \to 0$, a high SNR approximation of \eqref{OP Nearest B GC} with $1 - {e^{ - x}} \approx x$ is given by
\begin{align}\label{GC quadrature CDF B appox nearest}
{P_{{{\rm{B}}_{{{\rm{i}}^{\rm{*}}}}}}} \approx {b_1}{\varepsilon _{{{\rm{A}}_i}}}\sum\limits_{n = 1}^N {\left( {\sqrt {1 - {\phi _n}^2} \left( {1 + c_{n*}^\alpha } \right){c_{n*}}{e^{ - \pi {\lambda _{{\Phi _{\rm{B}}}}}c_{n*}^2}}} \right)}.
\end{align}
Substituting \eqref{GC quadrature CDF B appox nearest} into \eqref{Diversity random}, we obtain that the diversity gain for the near users of NNNF is one, which indicates that using NNNF will not affect the diversity gain.

\emph{Far users:}
For the far users, substituting \eqref{OP Nearest A GC} into \eqref{Diversity random}, we obtain that the diversity gain is still two. This indicates that NNNF will not affect the diversity gain.

\subsubsection{System Throughput in Delay-Sensitive Transmission Mode of NNNF}
Based on the analytical results for the outage probability of the near and far users, the system throughput of NNNF in the delay-sensitive transmission mode is given by
\begin{align}\label{throughput random}
{R_{{\tau _{{\rm{NNNF}}}}}} = \left( {1 - {P_{{{\rm{A}}_{{\rm{i^*}}}}}}} \right){R_1} + \left( {1 - {P_{{{\rm{B}}_{{\rm{i^*}}}}}}} \right){R_2},
\end{align}
where ${P_{{{\rm{A}}_{{\rm{i^*}}}}}}$ and ${P_{{{\rm{B}}_{{\rm{i^*}}}}}}$ are obtained from \eqref{OP Nearest A GC} and \eqref{OP Nearest B GC}, respectively.
\subsection{NNFF Selection Scheme}
In this scheme, we first select a user within disc ${D_{\rm{B}}}$ which has the shortest distance to the BS as a near NOMA user. Then we select a user within ring ${D_{\rm{A}}}$ which has the farthest distance to the BS as a far NOMA user (denoted by $\rm{A_{i'}}$). The use of this selection scheme is inspired by an interesting observation described in \cite{ding2014performance} that NOMA can offer a larger performance gain over conventional MA when user channel conditions are more distinct.
\subsubsection{Outage Probability of the Near Users of NNFF}
Since the same criterion for the near users is used, the outage probabilities of near nears for an arbitrary $\alpha$ and the special case $\alpha=2$ are the same as those expressed in \eqref{OP Nearest B GC} and \eqref{OP Nearest B GC alpha 2}, respectively.
\subsubsection{Outage Probability of the Far Users of NNFF}
Using the same definition of the outage probability of the far users, and similar to \eqref{OP A 1_1}, we can characterize the outage probability of the far users of NNFF.
The following theorem provides the outage probability of the far user of NNFF for an arbitrary choice of $\alpha$.
\begin{theorem}\label{theorem:5}
\emph{Conditioned on the PPPs and assuming ${R_{{D_{\rm{C}}}}} \gg {R_{{D_{\rm{B}}}}}$, the outage probability of $\mathrm{A_{i'}}$ can be approximated as follows:}
\begin{align}\label{OP farthest A GC}
&{P_{{{\rm{A}}_{{\rm{i'}}}}}} \approx {\varsigma ^*}\sum\limits_{n = 1}^N {\sqrt {1 - {\phi _n}^2} \left( {1 + c_{n*}^\alpha } \right){c_{n*}}{e^{ - \pi {\lambda _{{\Phi _{\rm{B}}}}}c_{n*}^2}}} \nonumber\\
&\times \sum\limits_{k = 1}^K {\sqrt {1 - \psi _k^2} {{\left( {1 + s_k^\alpha } \right)}^2}{s_k}{e^{ - \pi {\lambda _{{\Phi _{\rm{A}}}}}\left( {R_{{D_{\rm{A}}}}^2 - s_k^2} \right)}}} \sum\limits_{m = 1}^M {\sqrt {1 - \varphi _m^2} } \nonumber\\
&\times {e^{ - \left( {1 + s_k^\alpha } \right){t_m}}}{\chi _{{t_m}}}\left( {\ln \frac{{{\chi _{{t_m}}}\left( {1 + s_k^\alpha } \right)\left( {1 + c_{n*}^\alpha } \right)}}{{\eta \rho }} + 2{c_0}} \right) \nonumber\\
& + {b_3}{b_4}\sum\limits_{k = 1}^K {\sqrt {1 - {\psi _k}^2} (1 + s_k^\alpha ){s_k}{e^{\pi {\lambda _{{\Phi _{\rm{A}}}}}s_k^2}}} \nonumber\\
&\times\sum\limits_{n = 1}^N {\left( {\sqrt {1 - {\phi _n}^2} \left( {1 + c_{n*}^\alpha } \right){c_{n*}}{e^{ - \pi {\lambda _{{\Phi _{\rm{B}}}}}c_{n*}^2}}} \right)},
\end{align}
\emph{where} ${b_4} = \frac{{{\xi _{\rm{A}}}{e^{ - \pi {\lambda _{{\Phi _{\rm{A}}}}}R_{{D_{\rm{A}}}}^2}}{\omega _K}{\varepsilon _{{{\rm{A}}_i}}}}}{{{R_{{D_{\rm{A}}}}} + {R_{{D_{\rm{C}}}}}}}$.
\begin{proof}
See Appendix~D.
\end{proof}
\end{theorem}
\begin{corollary}\label{corollary:5}
\emph{For the special case} $\alpha=2$, \emph{after some manipulations, the outage probability of} $\mathrm{A_{i'}}$ \emph{can be simplified  as} \eqref{OP farthest A GC alpha 2} \emph{at the top of the next page}.
\begin{figure*}[!t]
\normalsize
\begin{align}\label{OP farthest A GC alpha 2}
&{\left. {{P_{{{\rm{A}}_{{\rm{i'}}}}}}} \right|_{\alpha  = 2}} \approx {\varsigma ^*}\sum\limits_{n = 1}^N {\sqrt {1 - {\phi _n}^2} \left( {1 + c_{n*}^2} \right){c_{n*}}{e^{ - \pi {\lambda _{{\Phi _{\rm{B}}}}}c_{n*}^2}}} \sum\limits_{k = 1}^K {\sqrt {1 - \psi _k^2} {{\left( {1 + s_k^2} \right)}^2}{s_k}{e^{ - \pi {\lambda _{{\Phi _{\rm{A}}}}}\left( {R_{{D_{\rm{A}}}}^2 - s_k^2} \right)}}} \nonumber\\
&\times \sum\limits_{m = 1}^M {\sqrt {1 - \varphi _m^2} \left( {{e^{ - \left( {1 + s_k^2} \right){t_m}}}{\chi _{{t_m}}}\left( {\ln \frac{{{\chi _{{t_m}}}\left( {1 + s_k^2} \right)\left( {1 + c_{n*}^2} \right)}}{{\eta \rho }} + 2{c_0}} \right)} \right)}  \nonumber\\
& + \frac{{{\xi _{\rm{A}}}{e^{ - \pi {\lambda _{{\Phi _{\rm{A}}}}}R_{{D_{\rm{A}}}}^2}}}}{2}\left( {\frac{{{e^{\pi {\lambda _{{\Phi _{\rm{A}}}}}R_{{D_{\rm{A}}}}^2}} - {e^{\pi {\lambda _{{\Phi _{\rm{A}}}}}R_{{D_{\rm{C}}}}^2}}}}{{\pi {\lambda _{{\Phi _{\rm{A}}}}}}} - \frac{{{e^{ - {\varepsilon _{{{\rm{A}}_i}}}}}}}{{\pi {\lambda _{{\Phi _{\rm{A}}}}} - {\varepsilon _{{{\rm{A}}_i}}}}}\left( {{e^{R_{{D_{\rm{A}}}}^2\left( {\pi {\lambda _{{\Phi _{\rm{A}}}}} - {\varepsilon _{{{\rm{A}}_i}}}} \right)}} - {e^{R_{{D_{\rm{C}}}}^2\left( {\pi {\lambda _{{\Phi _{\rm{A}}}}} - {\varepsilon _{{{\rm{A}}_i}}}} \right)}}} \right)} \right)\nonumber\\
&\times \frac{{{\xi _{\rm{B}}}}}{2}\left( {\frac{{\left( {{e^{ - R_{{D_{\rm{B}}}}^2\left( {\pi {\lambda _{{\Phi _{\rm{B}}}}} + {\varepsilon _{{{\rm{A}}_i}}}} \right) - {\varepsilon _{{{\rm{A}}_i}}}}} - {e^{ - {\varepsilon _{{{\rm{A}}_i}}}}}} \right)}}{{\left( {\pi {\lambda _{{\Phi _{\rm{B}}}}} + {\varepsilon _{{{\rm{A}}_i}}}} \right)}} - \frac{{\left( {{e^{ - \pi {\lambda _{{\Phi _{\rm{B}}}}}R_{{D_{\rm{B}}}}^2}} - 1} \right)}}{{\pi {\lambda _{{\Phi _{\rm{B}}}}}}}} \right).
\end{align}
\hrulefill \vspace*{0pt}
\end{figure*}
\end{corollary}

\subsubsection{Diversity Analysis}

Similarly, we provide diversity analysis of both the near and far uses in NNFF.

\emph{Near users:}
Since the same criterion for selecting a near user is used, the diversity gain is one, which is the same as for NNNF.

\emph{Far users:}
Substituting \eqref{OP farthest A GC} into \eqref{Diversity random}, we find that the diversity gain is still two. Therefore, we conclude that using opportunistic user selection schemes (NNNF and NNFF) based on distances will not affect the diversity gain.

\subsubsection{System Throughput in Delay-Sensitive Transmission Mode of NNFF}
Based on the analytical results for the outage probability of the near and far users, the system throughput of NNFF in the delay-sensitive transmission mode is given by
\begin{align}\label{throughput FFUS}
{R_{{\tau _{{\rm{NNFF}}}}}} = \left( {1 - {P_{{{\rm{A}}_{{\rm{i'}}}}}}} \right){R_1} + \left( {1 - {P_{{{\rm{B}}_{{\rm{i^*}}}}}}} \right){R_2},
\end{align}
where ${P_{{{\rm{A}}_{{\rm{i'}}}}}}$ and ${P_{{{\rm{B}}_{{\rm{i^*}}}}}}$ are obtained from \eqref{OP farthest A GC} and \eqref{OP Nearest B GC}, respectively.

\section{Numerical Results}\label{Numerical Results}
In this section, numerical results are presented to facilitate the performance evaluations (including the outage probability of the near and the far users and the delay sensitive throughput) of the proposed cooperative SWIPT NOMA protocol. In the considered network, we assume that the energy conversion efficiency of SWIPT is $\eta=0.7$ and the power allocation coefficients of NOMA is ${\left| {{p_{i1}}} \right|^2}=0.8$, ${\left| {{p_{i1}}} \right|^2}=0.2$. In the following figures, we use red, blue and black color lines to represent the RNRF, NNNF and NNFF user selection schemes, respectively.
\subsection{Outage Probability of the Near Users}
\begin{figure}[t!]
    \begin{center}
        \includegraphics[width=3.8 in]{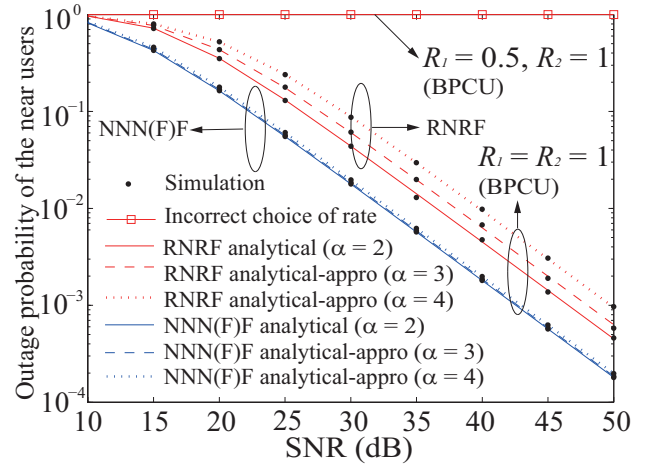}
        \caption{Outage probability of the near users versus SNR with different $\alpha$, where  $R_{{D_{\rm{B}}}}=2$ $m$, and ${\lambda _{{\Phi _{\rm{B}}}}}=1$.}
        \label{P_B_out_alpha}
    \end{center}
\end{figure}

\begin{figure}[t!]
    \begin{center}
        \includegraphics[width=3.8 in]{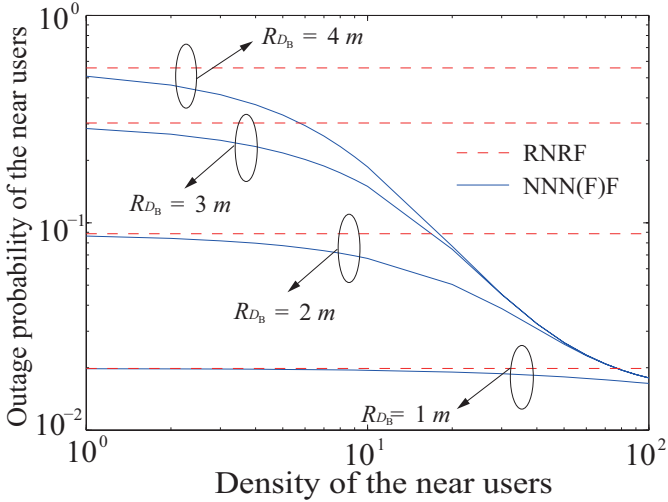}
        \caption{Outage probability of the near users versus density with different $R_{{D_{\rm{B}}}}$, where ${\lambda _{{\Phi _{\rm{B}}}}}=1$, and $SNR=30$ dB.}
        \label{P_B_out_density}
    \end{center}
\end{figure}

\begin{figure}[t!]
    \begin{center}
        \includegraphics[width=3.6 in]{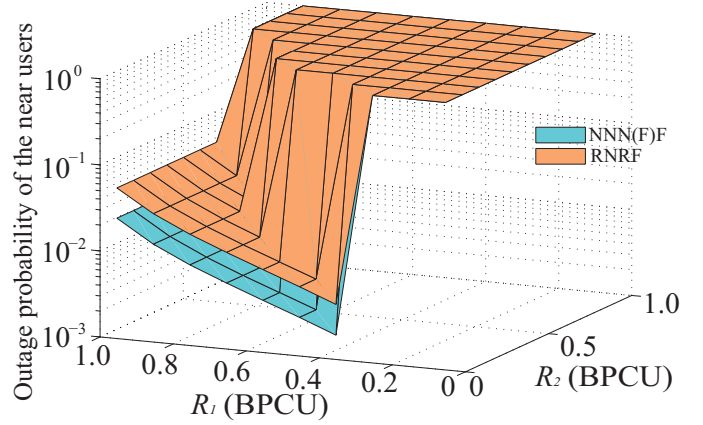}
        \caption{Outage probability of the near users versus $R_1$ and $R_2$, where $\alpha=2, R_{{D_{\rm{B}}}}=2$ $m$, and $SNR=30$ dB.}
        \label{P_B_out_rate1}
    \end{center}
\end{figure}

In this subsection, the outage probability achieved by the near users with different choices of density and path loss coefficients for the three user selection schemes is demonstrated. Note that the same user selection criterion is applied for the near users of NNNF and NNFF, we use NNN(F)F to represent these two selection schemes in Fig. \ref{P_B_out_alpha}, Fig. \ref{P_B_out_density}, and Fig. \ref{P_B_out_rate1}.

Fig. \ref{P_B_out_alpha} plots the outage probability of the near users versus SNR with different path loss coefficients for both RNRF and NNN(F)F.  The solid red and blue curves are for the special case $\alpha=2$ of RNRF and NNN(F)F, corresponding to the analytical results derived in \eqref{OP B alpha 2} and \eqref{OP Nearest B GC alpha 2}, respectively. The dashed red and blue curves are for an arbitrary choice of $\alpha$, corresponding to the analytical results derived in \eqref{OP B GC} and \eqref{OP Nearest B GC}, respectively. Monte Carlo simulation results are marked as ``$\bullet$" to verify our derivation. The figure shows precise agreement between the simulation and analytical curves. One can observe that by performing NNNF and NNFF (which we refer to as NNN(F)F in the figure), lower outage probability is achieved than with RNRF since shorter distances mean lower path loss and leads to better  performance. The figure also demonstrates that as $\alpha$ increases, outage will occur more frequently because of higher path loss. For NNNF and NNFF, the performance is very close for different values of $\alpha$. This is because we use the bounded path loss model (i.e. $1+d_{\rm{i}}^\alpha > 1$) to ensure that the path loss is always larger than one. When selecting the nearest near user, $d_{\rm{i}}$ will approach zero and the path loss will approach one, which makes the performance difference of the three selection schemes insignificant. It is worth noting that all   curves have the same slopes, which indicates that the diversity gains of the schemes  are the same. This phenomenon validates the insights we obtained from the analytical results   derived in \eqref{Diversity random}. Fig. \ref{P_B_out_alpha} also shows that if the choices of rates for users are incorrect (i.e.,  $ R_1=0.5$ and $R_2=1$ in this figure), the outage probability of the near users will be always one, which verifies the analytical results in \eqref{OP B GC} and \eqref{OP Nearest B GC}.

Fig. \ref{P_B_out_density} plots the outage probability of the near users versus their density with different values of $R_{{D_{\rm{B}}}}$. RNRF is also shown in the figure as a benchmark for comparison. Several observations are drawn as follows: 1) The outage probabilities of RNRF and NNN(F)F decrease with decreasing $R_{{D_{\rm{B}}}}$ because   path loss is reduced; 2) The outage probability of  NNN(F)F decreases as the density of the near users increases. This is due to  the multiuser diversity gain, since there is an increasing number of the near users; 3) The outage probability of RNRF is a constant, i.e., independent of the density of near users, and is the outage ceiling of the NNN(F)F. This is due to the fact that no opportunistic user selection is carried out for RNRF; and 4) An outage floor exits even if the density of the near users goes to infinity. This is due to the bounded path loss model we have used. When the number of the near users exceeds a threshold, the selected near user will be very close to the source, which  makes the path gain approach one.

Fig. \ref{P_B_out_rate1} plots the outage probability of the near users versus the rate of the near users and far users for both RNRF and NNN(F)F. One can observe that the outage of the near users occurs more frequently as the rate of the far user, $R_1$, increases. This is because in our proposed protocol, the near user $\mathrm{B_i}$ needs to first decode $x_{i1}$ which is intended to the far user $\mathrm{A_i}$, and then decode its own message. Therefore increasing $R_1$ makes it harder to decode $x_{i1}$, which will lead to increased outages. An important observation is that incorrect choices of $R_1$ and $R_2$ will make the outage probability  always   one. Particularly, for the choice of $R_1$, it should satisfy the condition ($ {\left| {{p_{i1}}} \right|^2} - {\left| {{p_{i2}}} \right|^2}{\tau _1} > 0$) in order to ensure that  successive interference cancelation  can be implemented. For the choice of $R_2$, it should satisfy the condition that the split energy for detecting $x_{i1}$ is also sufficient to detect $x_{i2}$ (${\varepsilon _{{{\rm{A}}_i}}} \ge {\varepsilon _{{{\rm{B}}_i}}}$).


\subsection{Outage Probability of the Far Users}
In this subsection, we demonstrate the outage probability of the far users  with different choices of the density, path loss coefficients, and user zone of the three user selection schemes.

\begin{figure}[t!]
    \begin{center}
        \includegraphics[width=3.8 in]{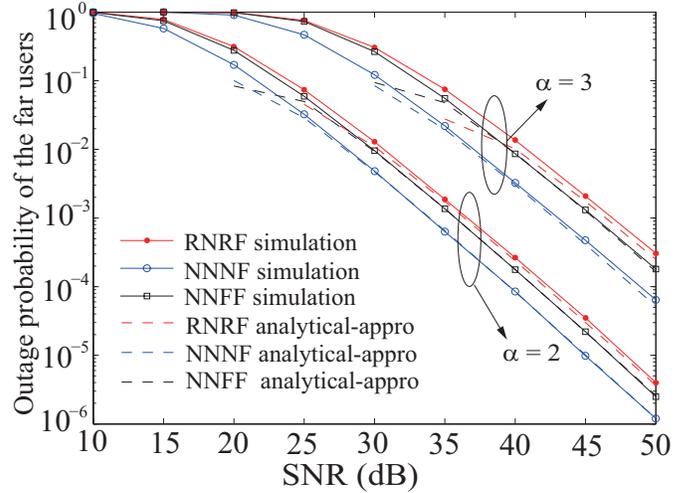}
        \caption{Outage probability of the far users with different $\alpha$, $ R_1=0.3$ BPCU, $R_{{D_{\rm{A}}}}=10$ $m$, $R_{{D_{\rm{B}}}}=2$ $m$, $R_{{D_{\rm{C}}}}=8$ $m$, ${\lambda _{{\Phi _{\rm{A}}}}}=1$, and ${\lambda _{{\Phi _{\rm{B}}}}}=1$.}
        \label{P_out_A_alpha}
    \end{center}
\end{figure}

Fig. \ref{P_out_A_alpha} plots the outage probability of the far users versus SNR with different path loss coefficients of RNRF, NNNF, and NNFF.  The dashed red, blue, and black curves circled together and pointed by $\alpha=2$, are the analytical approximations for the special case of RNRF, NNNF, and NNFF, which are obtained from \eqref{OP A GC alpha 2}, \eqref{OP Nearest A GC alpha 2} and \eqref{OP farthest A GC alpha 2}, respectively. The dashed red, blue, and black curves circled together and pointed by $\alpha=3$, are the analytical approximations for an arbitrary choice of $\alpha$ of RNRF, NNNF, and NNFF, which are obtained from \eqref{OP A GC}, \eqref{OP Nearest A GC} and \eqref{OP farthest A GC}, respectively. We use the solid marked lines to represent the Monte Carlo simulation results for each case. As can be observed from the figure, the simulation and the analytical approximation are very close, particularly in the high SNR region. Several observations can be drawn as follows: 1) NNNF achieves the lowest outage probability among the three selection schemes since both the near and far users have the smallest path loss; 2) NNFF achieves lower outage than RNRF, which indicates that the distance of the near users has more impact than that of the far users; 3) it is clear that all of the curves in Fig. \ref{P_out_A_alpha} have the same slopes, which indicates that the diversity gains of the far users for the three schemes are the same. In the diversity analysis, we showed that the diversity gain of the three selection schemes is two. The simulation validates the analytical results and indicates that the achievable diversity gain is the same for different user selection schemes.

\begin{figure}[t!]
    \begin{center}
        \includegraphics[width=3.8 in]{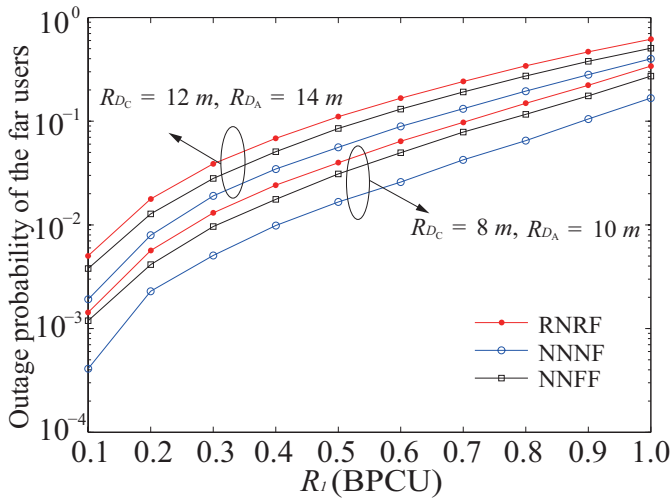}
        \caption{Outage probability of the far users versus $R_1$, where $\alpha=2$, $R_{{D_{\rm{B}}}}=2$ $m$, and $SNR=30$ dB.}
        \label{P_A_out_rate2}
    \end{center}
\end{figure}

Fig. \ref{P_A_out_rate2} plots the outage probability of the far users versus $R_1$ with different $R_{{D_{\rm{C}}}}$ and $R_{{D_{\rm{B}}}}$. One can observe   that the outage probabilities of the three schemes increase as   $R_1$ increases. This is because increasing $R_1$ will make the threshold of decoding   higher, which in turn leads to more outage. It can also be observed that increasing the radius of the user zone for the far users will deteriorate   the outage performance. The reason is that the path loss of the far users becomes larger.

\begin{figure}[t!]
    \begin{center}
        \includegraphics[width=3.8 in]{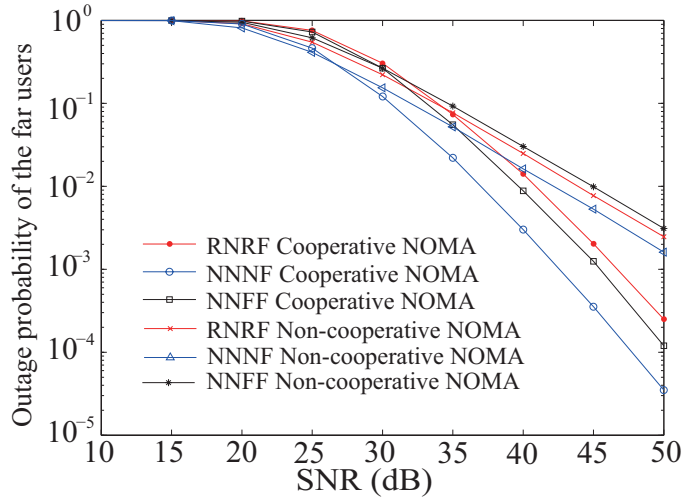}
        \caption{Comparison of outage probability with non-cooperative NOMA, $\alpha=3$, $R_1=0.3$ BPCU, $R_{{D_{\rm{A}}}}=10$ $m$, $R_{{D_{\rm{B}}}}=2$ $m$, $R_{{D_{\rm{C}}}}=8$ $m$, ${\lambda _{{\Phi _{\rm{A}}}}}=1$, and ${\lambda _{{\Phi _{\rm{B}}}}}=1 $.}
        \label{P_out_A_conventional}
    \end{center}
\end{figure}

Fig. \ref{P_out_A_conventional} plots the outage probability of the far users versus SNR for both cooperative NOMA and non-cooperative NOMA\footnote{It is common to use outage probability as a criterion to compare the performance of cooperative transmission and non-cooperative transmission schemes \cite{laneman2004cooperative}. In the context of cooperative NOMA, the use of outage probability is particularly useful  since the purpose of cooperative NOMA is to improve the reception reliability of the far users.}. Several observations can be drawn as follows: 1) by using an energy constrained relay to perform cooperative NOMA transmission, the outage probability of the far users has a larger slope than that of non-cooperative NOMA,  for all  user selection schemes. This is due to the fact that   cooperative NOMA can achieve a larger diversity gain and guarantees more reliable reception for the far users in the high SINR region; 2) NNNF achieves the lowest outage probability among these three selection schemes both for cooperative NOMA and non-cooperative NOMA because of its smallest path loss; 3) it is worth noting that NNFF has higher outage probability than RNRF in non-cooperative NOMA, however, it achieves lower outage probability than RNRF in cooperative NOMA. This phenomenon indicates that it is very helpful and necessary to apply cooperative NOMA in NNFF due to the largest performance gain over non-cooperative NOMA.

\subsection{Throughput in Delay-Sensitive Transmission Mode}

\begin{figure}[t!]
    \begin{center}
        \includegraphics[width=3.8 in]{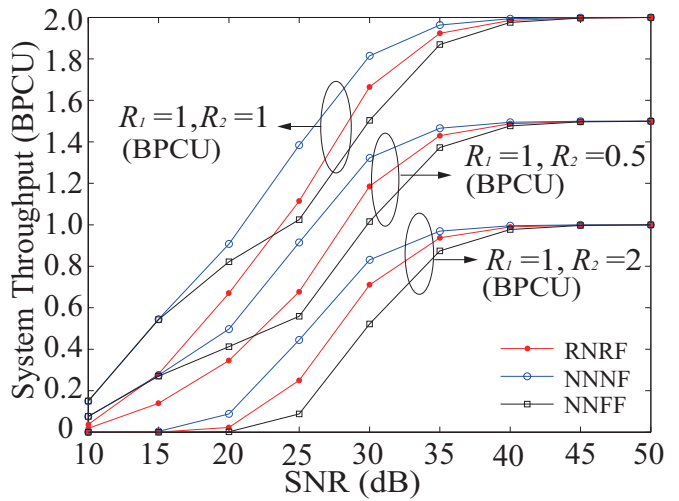}
        \caption{System throughput in delay-sensitive mode versus SNR with different rate, $\alpha=2$, $R_{{D_{\rm{A}}}}=10$ $m$, $R_{{D_{\rm{B}}}}=2$ $m$, $R_{{D_{\rm{C}}}}=8$ $m$, ${\lambda _{{\Phi _{\rm{A}}}}}=1$, and ${\lambda _{{\Phi _{\rm{B}}}}}=1 $.}
        \label{Throughput}
    \end{center}
\end{figure}

Fig. \ref{Throughput} plots the system throughput versus SNR with different targeted  rates. One can observe that NNNF achieves the highest throughput since it has the lowest outage probability among   three selection schemes. The figure also demonstrates the existence of the  throughput ceilings  in the high SNR region. This is due to the fact that the outage probability is approaching zero and the throughput is  determined only by the targeted data  rate. It is worth noting that increasing $R_2$ from $R_2=0.5$ BPCU to $R_2=1$ BPCU can improve the throughput; however, for the case $R_2=2$ BPCU, the throughput is lowered. This is because, in the latter case, the energy remaining for information decoding is  not sufficient for message detection of the near user, and hence an outage occurs, which in turn affects the throughput. Therefore, we see that it is important  to select appropriate transmission rates when designing  practical NOMA downlink transmission systems.

\section{Conclusions}\label{Conclusions}
In this paper, the application of SWIPT to NOMA has been considered. A novel cooperative SWIPT NOMA protocol with three different user selection criteria  has been proposed. We have used the stochastic geometric approach to provide a complete framework to model the locations of users and evaluate the performance of the proposed user selection schemes. Closed-form results have been derived in terms of outage probability and delay-sensitive throughput to determine the system performance. The diversity gain of the three user selection schemes has also been characterized and proved to be the same as that of a conventional cooperative network. For the proposed protocol, the decreasing rate of the outage probability of far users is $\frac{{\ln SNR}}{{SN{R^2}}}$ while it is $\frac{1}{{SN{R^2}}}$ for a conventional cooperative network. Numerical results have been presented to validate our analysis. We conclude that by carefully choosing the parameters of the network, (e.g., transmission rate or power splitting coefficient), acceptable system performance can be guaranteed even if the users do not use their own batteries  to power the relay transmission.
\numberwithin{equation}{section}
\section*{Appendix~A: Proof of Theorem~\ref{theorem:2}} \label{Appendix:A}
\renewcommand{\theequation}{A.\arabic{equation}}
\setcounter{equation}{0}
Substituting \eqref{DT SNR B x1} and \eqref{A SNR total} into \eqref{OP A 1_1}, the outage probability can be expressed as follows:
\begin{align}\label{OP A 1}
{P_{{{\rm{A_i}}}}} =&\underbrace {\Pr \left( {\gamma _{{{\rm{A}}_{\rm{i}}},{\rm{MRC}}}^{{x_{i1}}} < {\tau _1},\frac{{\rho {{\left| {{h_{{{\rm{B}}_{\rm{i}}}}}} \right|}^2}{{\left| {{p_{i1}}} \right|}^2}}}{{\rho {{\left| {{h_{{{\rm{B}}_{\rm{i}}}}}} \right|}^2}{{\left| {{p_{i2}}} \right|}^2} + 1 + d_{{{\rm{B}}_{\rm{i}}}}^\alpha }} > {\tau _1}} \right)}_{{\Theta _1}}\nonumber\\
& + \underbrace {\Pr \left( {\gamma _{{\rm{S}},{{\rm{A}}_{\rm{i}}}}^{{x_{i1}}} < {\tau _1},\frac{{\rho {{\left| {{h_{{{\rm{B}}_{\rm{i}}}}}} \right|}^2}{{\left| {{p_{i1}}} \right|}^2}}}{{\rho {{\left| {{h_{{{\rm{B}}_{\rm{i}}}}}} \right|}^2}{{\left| {{p_{i2}}} \right|}^2} + 1 + d_{{{\rm{B}}_{\rm{i}}}}^\alpha }} < {\tau _1}} \right)}_{{\Theta _2}},
\end{align}

We express ${\Theta _1}$  as \eqref{Theta 1} on the top of next page
\begin{figure*}[!t]
\normalsize
\begin{align}\label{Theta 1}
&{\Theta _1} = \Pr \left( {{Z_i} < \frac{{{\tau _1} - \frac{{\rho {X_i}{{\left| {{p_{i1}}} \right|}^2}}}{{\rho {X_i}{{\left| {{p_{i2}}} \right|}^2} + 1}}}}{{\eta \rho \left( {{Y_i} - {\varepsilon _{{{\rm{A}}_i}}}} \right)}},{X_i} < {\varepsilon _{{{\rm{A}}_i}}},{Y_i} > {\varepsilon _{{{\rm{A}}_i}}}} \right)\nonumber\\
= & \int\limits_{{{D_{{{\rm{B}}}}}}} {\int\limits_{{D_{{{\rm{A}}}}}}{\int_0^{{\varepsilon _{{{\rm{A}}_i}}}} {\underbrace {\int_{{\varepsilon _{{{\rm{A}}_i}}}}^\infty  {\left( {1 - {e^{ - \left( {1 + d_{{{\rm{C}}_{\rm{i}}}}^\alpha } \right)\frac{{{\tau _1} - \frac{{\rho x{{\left| {{p_{i1}}} \right|}^2}}}{{\rho x{{\left| {{p_{i2}}} \right|}^2} + 1}}}}{{\eta \rho \left( {y - {\varepsilon _{{{\rm{A}}_i}}}} \right)}}}}} \right){f_{Y_i}}\left( y \right)dy} }_\Xi } } } {f_{X_i}}\left( x \right)dx{f_{{W_{{{\rm{A}}_{\rm{i}}}}}}}\left( {{\omega _{{{\rm{A_i}}}}}} \right)d{\omega _{{{\rm{A_i}}}}}{f_{{W_{{{\rm{B}}_{\rm{i}}}}}}}\left( {{\omega _{{{\rm{B_i}}}}}} \right)d{\omega _{{{\rm{B_i}}}}},
\end{align}
\hrulefill \vspace*{0pt}
\end{figure*}
where ${f_{X_i}}\left( x \right) = \left( {1 + d_{{{\rm{A}}_{\rm{i}}}}^\alpha } \right){e^{ - \left( {1 + d_{{{\rm{A}}_{\rm{i}}}}^\alpha } \right)x}}$, and ${f_{Y_i}}\left( y \right) = \left( {1 + d_{{{\rm{B}}_{\rm{i}}}}^\alpha } \right){e^{ - \left( {1 + d_{{{\rm{B}}_{\rm{i}}}}^\alpha } \right)y}}$.

Based on \eqref{Theta 1}, using $t = y - {\varepsilon _{{{\rm{A}}_i}}}$, we calculate ${\Xi}$ as follows:
\begin{align}\label{Xi 2}
{\Xi} =& \int_0^\infty  {\left( {1 - {e^{ - \left( {1 + d_{{{\rm{C}}_{\rm{i}}}}^\alpha } \right)\frac{{\tau  - \frac{{\rho x{{\left| {{p_{i1}}} \right|}^2}}}{{\rho x{{\left| {{p_{i2}}} \right|}^2} + 1}}}}{{\eta \rho t}}}}} \right)} \nonumber\\
&\times\left( {1 + d_{{{\rm{B}}_{\rm{i}}}}^\alpha } \right){e^{ - \left( {1 + d_{{{\rm{B}}_{\rm{i}}}}^\alpha } \right)\left( {t + {\varepsilon _{{{\rm{A}}_i}}}} \right)}}dt.
\end{align}
Applying \cite[ Eq. (3.324)]{gradshteyn}, we rewrite \eqref{Xi 2} as follows:
\begin{align}\label{Xi 2 1}
\Xi  = {e^{ - \left( {1 + d_{{{\rm{B}}_{\rm{i}}}}^\alpha } \right){\varepsilon _{{{\rm{A}}_i}}}}}\left( {1 - 2\sqrt {\chi \Lambda } {K_1}\left( {2\sqrt {\chi \Lambda } } \right)} \right),
\end{align}
where $\Lambda  = \frac{{\left( {1 + d_{{{\rm{B}}_{\rm{i}}}}^\alpha } \right)\left( {1 + d_{{{\rm{C}}_{\rm{i}}}}^\alpha } \right)}}{{\eta \rho }}$ and $\chi  = {\tau _1} - \frac{{\rho x{{\left| {{p_{i1}}} \right|}^2}}}{{\rho x{{\left| {{p_{i2}}} \right|}^2} + 1}}$.

We use the series representation of Bessel functions to obtain the high SNR approximation which is expressed as follows:
\begin{align}\label{Bessel appro}
x{K_1}\left( x \right) \approx 1 + \frac{{{x^2}}}{2}\left( {\ln \frac{x}{2} + {c_0}} \right),
\end{align}
where ${K_1}\left(  \cdot  \right)$ is the modified Bessel function for the seconde kind, ${c_0} =  - \frac{{\varphi \left( 1 \right)}}{2} - \frac{{\varphi \left( 2 \right)}}{2}$, and $\varphi \left(  \cdot  \right)$ denotes the psi function \cite{gradshteyn}.

To obtain the high SNR approximation of \eqref{Xi 2 1} and using \eqref{Bessel appro}, we obtain
\begin{align}\label{Xi 2 Bessel appro}
\Xi  \approx  - \chi \Lambda \left( {\ln \chi \Lambda  + 2{c_0}} \right).
\end{align}
Substituting \eqref{Xi 2 Bessel appro} into \eqref{Theta 1}, we rewrite \eqref{Theta 1} as follows:
\begin{align}\label{Theta 1_2}
{\Theta _1} =&  - \int_0^{{\varepsilon _{{{\rm{A}}_i}}}} \chi  \int\limits_{{D_{{{\rm{A}}_{\rm{i}}}}}} {\Lambda {e^{ - \left( {1 + d_{{{\rm{A}}_{\rm{i}}}}^\alpha } \right)x}}} \nonumber\\
&\times\underbrace {\int\limits_{{D_{{{\rm{B}}_{\rm{i}}}}}} {\left( {1 + d_{{{\rm{B}}_{\rm{i}}}}^\alpha } \right)\left( {\ln \chi \Lambda  + 2{c_0}} \right){f_{{W_{{{\rm{B}}_{\rm{i}}}}}}}\left( {{\omega _{{{\rm{B}}_{\rm{i}}}}}} \right)d{\omega _{{{\rm{B}}_{\rm{i}}}}}} }_\Phi\nonumber\\
&{f_{{W_{{{\rm{A}}_{\rm{i}}}}}}}\left( {{\omega _{{{\rm{A}}_{\rm{i}}}}}} \right)d{\omega _{{{\rm{A}}_{\rm{i}}}}}dx.
\end{align}

Since ${d_{{{\rm{C}}_{\rm{i}}}}} = \sqrt {d_{{{\rm{A}}_{\rm{i}}}}^2 + d_{{{\rm{B}}_{\rm{i}}}}^2 - 2{d_{{{\rm{A}}_{\rm{i}}}}}{d_{{{\rm{B}}_{\rm{i}}}}}\cos \left( {{\theta _i}} \right)}$ and ${R_{{D_{\rm{C}}}}} \gg {R_{{D_{\rm{B}}}}}$, we can approximate the distance as ${d_{{{\rm{A}}_{\rm{i}}}}} \approx {d_{{{\rm{C}}_{\rm{i}}}}}$. Applying \eqref{W_B}, we calculate $\Phi$ as follows:
\begin{align}\label{Phi}
\Phi  \approx& \frac{2}{{R_{{D_{{{\rm{B}}}}}}^2}}\int_0^{{R_{{D_{\rm{B}}}}}} {\left( {1 + {r^\alpha }} \right)\left( {\ln {m_0}\left( {1 + {r^\alpha }} \right) + 2{c_0}} \right)rdr},
\end{align}
where ${m_0} = \frac{{\chi \left( {1 + d_{{{\rm{C}}_{\rm{i}}}}^\alpha } \right)}}{{\eta \rho }} \approx \frac{{\chi \left( {1 + d_{{{\rm{A}}_{\rm{i}}}}^\alpha } \right)}}{{\eta \rho }}$.

For an arbitrary choice of $\alpha$, the integral in \eqref{Phi} is mathematically intractable, we use Gaussian-Chebyshev quadrature to find the approximation. Then $\Phi$ can be approximated as follows:
\begin{align}\label{Phi_GC}
\Phi  \approx \frac{{{\omega _N}}}{2}\sum\limits_{n = 1}^N {\left( {\sqrt {1 - {\phi _n}^2} {c_n}\left( {\ln {m_0}{c_n} + 2{c_0}} \right)\left( {{\phi _n} + 1} \right)} \right)} .
\end{align}

Substituting \eqref{Phi_GC} into \eqref{Theta 1_2}, we rewrite \eqref{Theta 1_2} as follows:
\begin{align}\label{Theta 1_3}
&\Theta_1
= - \frac{{{\omega _N}}}{{R_{{D_{\rm{A}}}}^2 - R_{{D_{\rm{C}}}}^2}}\int_0^{{\varepsilon _{{{\rm{A}}_i}}}} {\frac{\chi }{{\eta \rho }}} \sum\limits_{n = 1}^N {\left( {{\phi _n} + 1} \right)\sqrt {1 - {\phi _n}^2} }   \nonumber\\
&\underbrace {\int_{{R_{{D_{\rm{C}}}}}}^{{R_{{D_{\rm{A}}}}}} {r{{\left( {1 + {r^\alpha }} \right)}^2}{e^{ - \left( {1 + {r^\alpha }} \right)x}}{c_n}\left( {\ln \frac{{\chi \left( {1 + {r^\alpha }} \right)}}{{\eta \rho }}{c_n} + 2{c_0}} \right)} dr}_\Delta \nonumber\\
&dx.
\end{align}
Similarly as above, we use Gaussian-Chebyshev quadrature to find an approximation of $\Delta$ in \eqref{Theta 1_3} as follows:
\begin{align}\label{Theta 1_3 GC}
\Delta  \approx &\frac{{{\omega _K}\left( {{R_{{D_{\rm{A}}}}} - {R_{{D_{\rm{C}}}}}} \right)}}{2}\sum\limits_{k = 1}^K {\sqrt {1 - \psi _k^2} {s_k}{{\left( {1 + s_k^\alpha } \right)}^2}} \nonumber\\
&\times{e^{ - \left( {1 + s_k^\alpha } \right)x}}{c_n}\left( {\ln \frac{{\chi \left( {1 + s_k^\alpha } \right)}}{{\eta \rho }}{c_n} + 2{c_0}} \right).
\end{align}

Substituting \eqref{Theta 1_3 GC} into \eqref{Theta 1_3}, we rewrite \eqref{Theta 1_3} as follows:
\begin{align}\label{Theta 1_4}
\Theta_1 =& {a_2}\sum\limits_{n = 1}^N {\left( {{\phi _n} + 1} \right)\sqrt {1 - {\phi _n}^2} {c_n}\sum\limits_{k = 1}^K {\sqrt {1 - \psi _k^2} {s_k}{{\left( {1 + s_k^\alpha } \right)}^2}} }  \nonumber\\
&\times \underbrace {\int_0^{{\varepsilon _{{{\rm{A}}_i}}}} {\chi {e^{ - \left( {1 + s_k^\alpha } \right)x}}\left( {\ln \frac{{\chi \left( {1 + s_k^\alpha } \right)}}{{\eta \rho }}{c_n} + 2{c_0}} \right)dx} }_\Psi,
\end{align}
where ${a_2} =-\frac{{{\omega _N}{\omega _K}}}{{2\left( {{R_{{D_{\rm{A}}}}} + {R_{{D_{\rm{C}}}}}} \right)\eta \rho }}$.

Similarly, we use Gaussian-Chebyshev quadrature to find an approximation of $\Psi$ in \eqref{Theta 1_4} as follows:
\begin{align}\label{Theta 1_4_GC}
\Psi  \approx& \frac{{{\omega _M}{\varepsilon _{{{\rm{A}}_i}}}}}{2}\sum\limits_{m = 1}^M {\sqrt {1 - \varphi _m^2} {e^{ - \left( {1 + s_k^\alpha } \right){t_m}}}}\nonumber\\
&\times{\chi _{{t_m}}}\left( {\ln \frac{{{\chi _{{t_m}}}\left( {1 + s_k^\alpha } \right)}}{{\eta \rho }}{c_n} + 2{c_0}} \right).
\end{align}

Substituting \eqref{Theta 1_4_GC} into \eqref{Theta 1_4}, we obtain
\begin{align}\label{Theta 1_5}
&\Theta_1 \approx \zeta_1 \sum\limits_{n = 1}^N {\left( {{\phi _n} + 1} \right)\sqrt {1 - {\phi _n}^2} } {c_n}\sum\limits_{k = 1}^K {\sqrt {1 - \psi _k^2} {s_k}{{\left( {1 + s_k^\alpha } \right)}^2}}\nonumber\\
&\sum\limits_{m = 1}^M {\sqrt {1 - \varphi _m^2} {e^{ - \left( {1 + s_k^\alpha } \right){t_m}}}{\chi _{{t_m}}}\left( {\ln \frac{{{\chi _{{t_m}}}\left( {1 + s_k^\alpha } \right)}}{{\eta \rho }}{c_n} + 2{c_0}} \right)}.
\end{align}

We express ${\Theta _2}$  as follows:
\begin{align}\label{Theta 2}
{\Theta _2}
=&\Pr \left( {{X_i} < {\varepsilon _{{{\rm{A}}_i}}}} \right)\Pr \left( {{Y_i} < {\varepsilon _{{{\rm{A}}_i}}}} \right).
\end{align}

The CDF of $X_i$ for ${\rm{A_i}}$ is given by
\begin{align}\label{Theta 2 CDF A}
{F_{X_i}}\left( \varepsilon  \right) &= \int\limits_D {\left( {1 - {e^{ - \left( {1 + d_{{{\rm{A}}_{\rm{i}}}}^\alpha } \right)\varepsilon }}} \right){f_{{W_{{{\rm{A}}_{\rm{i}}}}}}}\left( {{\omega _{{{\rm{A}}_{\rm{i}}}}}} \right)d{\omega _{{{\rm{A}}_{\rm{i}}}}}} \nonumber\\
& = \frac{2}{{R_{{D_{\rm{A}}}}^2 - R_{{D_{\rm{C}}}}^2}}\int_{{R_{{D_{\rm{C}}}}}}^{{R_{{D_{\rm{A}}}}}} {\left( {1 - {e^{ - \left( {1 + {r^\alpha }} \right)\varepsilon }}} \right)rdr}.
\end{align}

For an arbitrary choice of $\alpha$, similarly to \eqref{GC quadrature CDF B}, we provide Gaussian-Chebyshev quadrature to find the approximation for the CDF of $X_i$. We rewrite \eqref{Theta 2 CDF A} as follows:
\begin{align}\label{GC quadrature CDF A}
{F_{X_i}}\left( \varepsilon  \right) \approx \frac{{{\omega _K}}}{{{R_{{D_{\rm{A}}}}} + {R_{{D_{\rm{C}}}}}}}\sum\limits_{k = 1}^K {\sqrt {1 - {\psi _k}^2} \left( {1 - {e^{ - (1 + s_k^\alpha )\varepsilon }}} \right){s_k}}.
\end{align}

When $\varepsilon  \to 0$, a high SNR approximation of the \eqref{GC quadrature CDF A} is given by
\begin{align}\label{GC quadrature CDF A appro}
{F_{X_i}}\left( \varepsilon  \right) \approx \frac{{{\omega _K}\varepsilon }}{{{R_{{D_{\rm{A}}}}} + {R_{{D_{\rm{C}}}}}}}\sum\limits_{k = 1}^K {\sqrt {1 - {\psi _k}^2} (1 + s_k^\alpha ){s_k}}.
\end{align}

Substituting \eqref{GC quadrature CDF A appro} and \eqref{GC quadrature CDF B} into \eqref{Theta 2}, we can obtain the approximation for the general case as follows:
\begin{align}\label{Theta 2 GC quadrature}
{\Theta _2} \approx {a_1}\sum\limits_{n = 1}^N {\sqrt {1 - {\phi _n}^2} {c_n}\left( {{\phi _n} + 1} \right)} \sum\limits_{k = 1}^K {\sqrt {1 - {\psi _k}^2} (1 + s_k^\alpha ){s_k}}.
\end{align}

Combining \eqref{Theta 1_5} and \eqref{Theta 2 GC quadrature}, we can obtain \eqref{OP A GC}.

The proof is completed.

\section*{Appendix~B: Proof of Corollary~\ref{corollary:2}} \label{Appendix:B}
\renewcommand{\theequation}{B.\arabic{equation}}
\setcounter{equation}{0}
For the special case $\alpha  = 2$, and let $\lambda  =\left( {1 + {r^2}} \right)$, we rewrite~\eqref{Phi} as follows:
\begin{align}\label{Phi_1 alpha 2}
{\left. \Phi  \right|_{\alpha  = 2}}  =&\frac{1}{{R_{{D_{\rm{B}}}}^2}}\int_1^{1 + R_{{D_{\rm{B}}}}^2} {\lambda \left( {\ln {m_{{0^*}}}\lambda  + 2{c_0}} \right)d\lambda } \nonumber\\
=&\frac{{\left( {R_{{D_{\rm{B}}}}^2 + 2} \right)\ln {m_{{0^*}}}}}{2} + {b_0},
\end{align}
where ${{m_{{0^*}}}} = \frac{{\chi \left( {1 + d_{{{\rm{C}}_{\rm{i}}}}^2 } \right)}}{{\eta \rho }} \approx \frac{{\chi \left( {1 + d_{{{\rm{A}}_{\rm{i}}}}^2 } \right)}}{{\eta \rho }}$.

Substituting \eqref{Phi_1 alpha 2} and applying $\alpha=2$ into \eqref{Theta 1}, we obtain
\begin{align}\label{theta 1 alpha 2_1}
&{\left. {{\Theta _1}} \right|_{\alpha  = 2}} =  - \frac{{\left( {R_{{D_{\rm{B}}}}^2 + 2} \right)}}{{2\left( {R_{{D_{\rm{A}}}}^2 - R_{{D_{\rm{C}}}}^2} \right)\eta \rho }}\int_0^{{\varepsilon _{{{\rm{A}}_i}}}} \chi \nonumber\\
&\underbrace {\int_{{R_{{D_{\rm{C}}}}}}^{{R_{{D_{\rm{A}}}}}} {r{{\left( {1 + {r^2}} \right)}^2}{e^{ - \left( {1 + {r^2}} \right)x}}\left( {\ln \frac{{\chi \left( {1 + {r^2}} \right)}}{{\eta \rho }} + {b_0}} \right)dr} }_{{{\left. \Delta  \right|}_{\alpha  = 2}}}dx.
\end{align}

We notice that the integral ${\left. \Delta  \right|_{\alpha  = 2}}$ in \eqref{theta 1 alpha 2_1} is mathematically intractable. We use Gaussian-Chebyshev quadrature to find an approximation. Then ${\left. \Delta  \right|_{\alpha  = 2}}$  can be approximated as follows:
\begin{align}\label{delta alpha 2}
&{\left. \Delta  \right|_{\alpha  = 2}} \approx \frac{{{\omega _K}\left( {{R_{{D_{\rm{A}}}}} - {R_{{D_{\rm{C}}}}}} \right)}}{2}\nonumber\\
&\sum\limits_{k = 1}^K {\sqrt {1 - \psi _k^2} {s_k}{{\left( {1 + s_k^2} \right)}^2}{e^{ - \left( {1 + s_k^2} \right)x}}\left( {\ln \frac{{\chi \left( {1 + s_k^2} \right)}}{{\eta \rho }} + {b_0}} \right)}.
\end{align}

Substituting \eqref{delta alpha 2} into \eqref{theta 1 alpha 2_1}, we rewrite \eqref{theta 1 alpha 2_1} as follows:
\begin{align}\label{theta 1 alpha 2_2}
&{\left. {{\Theta _1}} \right|_{\alpha  = 2}} =  - \frac{{{\omega _K}\left( {R_{{D_{\rm{B}}}}^2 + 2} \right)}}{{4\left( {{R_{{D_{\rm{A}}}}} + {R_{{D_{\rm{C}}}}}} \right)\eta \rho }}\sum\limits_{k = 1}^K {\sqrt {1 - \psi _k^2} {s_k}{{\left( {1 + s_k^2} \right)}^2}} \nonumber\\
&\times\underbrace {\int_0^{{\varepsilon _{{{\rm{A}}_i}}}} {\chi {e^{ - \left( {1 + s_k^2} \right)x}}\left( {\ln \frac{{\chi \left( {1 + s_k^2} \right)}}{{\eta \rho }} + {b_0}} \right)dx} }_{{{\left. \Psi  \right|}_{\alpha  = 2}}}.
\end{align}

Similarly, we use Gaussian-Chebyshev quadrature to find the approximation of ${{{\left. \Psi  \right|}_{\alpha  = 2}}}$ in \eqref{theta 1 alpha 2_2} as follows:
\begin{align}\label{psi alpha 2_2_GC}
&{\left. \Psi  \right|_{\alpha  = 2}} \approx \frac{{{\omega _M}{\varepsilon _{{{\rm{A}}_i}}}}}{2}\sum\limits_{m = 1}^M {\sqrt {1 - \varphi _m^2} } \nonumber\\
&\times{\chi _{{t_m}}}{e^{ - \left( {1 + s_k^2} \right){t_m}}}\left( {\ln \frac{{{\chi _{{t_m}}}\left( {1 + s_k^2} \right)}}{{\eta \rho }}{c_n} + {b_0}} \right).
\end{align}

Substituting \eqref{psi alpha 2_2_GC} into \eqref{theta 1 alpha 2_2}, we obtain
\begin{align}\label{theta 1 alpha 2_3}
&{\left. {{\Theta _1}} \right|_{\alpha  = 2}} =  {\zeta _2}\sum\limits_{k = 1}^K {\sqrt {1 - \psi _k^2} {s_k}{{\left( {1 + s_k^2} \right)}^2}} \sum\limits_{m = 1}^M {\sqrt {1 - \varphi _m^2} }  \nonumber\\
&\times{\chi _{{t_m}}}{e^{ - \left( {1 + s_k^2} \right){t_m}}}\left( {\ln \frac{{{\chi _{{t_m}}}\left( {1 + s_k^2} \right)}}{{\eta \rho }}{c_n} + {b_0}} \right).
\end{align}

For the special case $\alpha=2$, the CDF of ${X_i}$ in \eqref{Theta 2 CDF A} can be calculated as follows:
\begin{align}\label{CDF A alpha 2}
{\left. {{F_{{X_i}}}\left( \varepsilon  \right)} \right|_{\alpha  = 2}} = 1 - \frac{{{e^{ - \left( {1 + R_{{D_{\rm{C}}}}^2} \right)\varepsilon }}}}{{\varepsilon \left( {R_{{D_{\rm{A}}}}^2 - R_{{D_{\rm{C}}}}^2} \right)}} + \frac{{{e^{ - \left( {1 + R_{{D_{\rm{A}}}}^2} \right)\varepsilon }}}}{{\varepsilon \left( {R_{{D_{\rm{A}}}}^2 - R_{{D_{\rm{C}}}}^2} \right)}}.
\end{align}

Substituting \eqref{CDF A alpha 2} and \eqref{CDF B alpha 2} into \eqref{Theta 2}, we can obtain $\Theta _2$ for the special case $\alpha=2$ in exact closed-form as follows:
\begin{align}\label{Theta 2 alhpa 2}
{\left. {{\Theta _2}} \right|_{\alpha  = 2}} =& \left( {1 - \frac{{{e^{ - \left( {1 + R_{{D_{\rm{C}}}}^2} \right){\varepsilon _{{{\rm{A}}_i}}}}}}}{{{\varepsilon _{{{\rm{A}}_i}}}\left( {R_{{D_{\rm{A}}}}^2 - R_{{D_{\rm{C}}}}^2} \right)}} + \frac{{{e^{ - \left( {1 + R_{{D_{\rm{A}}}}^2} \right){\varepsilon _{{{\rm{A}}_i}}}}}}}{{{\varepsilon _{{{\rm{A}}_i}}}\left( {R_{{D_{\rm{A}}}}^2 - R_{{D_{\rm{C}}}}^2} \right)}}} \right)\nonumber\\
&\times\left( {1 - \frac{{{e^{ - {\varepsilon _{{{\rm{A}}_i}}}}}}}{{R_{{D_{\rm{B}}}}^2{\varepsilon _{{{\rm{A}}_i}}}}} + \frac{{{e^{ - \left( {1 + R_{{D_{\rm{B}}}}^2} \right){\varepsilon _{{{\rm{A}}_i}}}}}}}{{R_{{D_{\rm{B}}}}^2{\varepsilon _{{{\rm{A}}_i}}}}}} \right).
\end{align}
Combining \eqref{theta 1 alpha 2_3} and \eqref{Theta 2 alhpa 2}, we can obtain \eqref{OP A GC alpha 2}.

The proof is completed.

\section*{Appendix~C: Proof of Theorem~\ref{theorem:4}} \label{Appendix:C}
\renewcommand{\theequation}{C.\arabic{equation}}
\setcounter{equation}{0}
Conditioned on the event that the numbers of users in group $\{{{{\rm{A}}_{{\rm{i}}}}}\}$ and $\{{{{\rm{B}}_{{\rm{i}}}}}\}$   satisfy ${\rm V} = {N_{\rm{A}}} \ge 1,{N_{\rm{B}}} \ge 1$, we express the outage probability for ${{{\rm{A}}_{{\rm{i*}}}}}$ by applying ${X_{{i^*}}} \to {X_i}$, ${Y_{{i^*}}} \to {Y_i}$, and ${Z_{{i^*}}} \to {Z_i}$ in \eqref{OP A 1} then obtain
\begin{align}\label{OP nearest A 1}
&{P_{{{\rm{A}}_{{{\rm{i}}^{\rm{*}}}}}}} = \underbrace {\Pr \left( {\left. {\frac{{\rho {X_{{i^*}}}{{\left| {{p_{i1}}} \right|}^2}}}{{\rho {X_{{i^*}}}{{\left| {{p_{i2}}} \right|}^2} + 1}} < {\tau _1},\frac{{\rho {Y_{{i^*}}}{{\left| {{p_{i1}}} \right|}^2}}}{{\rho {{\left| {{p_{i2}}} \right|}^2}{Y_{{i^*}}} + 1}} < {\tau _1}} \right|{\rm V}} \right)}_{\Theta _2^*}\nonumber\\
&+\underbrace {\Pr \left( {\left. {{Z_{{i^*}}} < \frac{{{\tau _1} - \frac{{\rho {X_{{i^*}}}{{\left| {{p_{i1}}} \right|}^2}}}{{\rho {X_{{i^*}}}{{\left| {{p_{i2}}} \right|}^2} + 1}}}}{{\eta \rho \left( {{Y_{{i^*}}} - {\varepsilon _{{{\rm{A}}_i}}}} \right)}},{X_{{i^*}}} < {\varepsilon _{{{\rm{A}}_i}}},{Y_{{i^*}}} > {\varepsilon _{{{\rm{A}}_i}}}} \right|{\rm V}} \right)}_{\Theta _1^*},
\end{align}
where ${X_{{i^*}}} = \frac{{{{\left| {{h_{{{\rm{A}}_{\rm{i}}}}}} \right|}^2}}}{{1 + d_{{{\rm{A}}_{{\rm{i^*}}}}}^\alpha }}$, ${Y_{{i^*}}} = \frac{{{{\left| {{h_{{{\rm{B}}_{\rm{i}}}}}} \right|}^2}}}{{1 + d_{{{\rm{B}}_{{\rm{i^*}}}}}^\alpha }}$, and ${Z_{{i^*}}} = \frac{{{{\left| {{g_i}} \right|}^2}}}{{1 + d_{{{\rm{C}}_{{\rm{i^*}}}}}^\alpha }}$. Here ${d_{{{\rm{A}}_{{\rm{i^*}}}}}}$, ${d_{{{\rm{B}}_{{\rm{i^*}}}}}}$, and ${d_{{{\rm{C}}_{{\rm{i^*}}}}}}$ are distances from the BS to ${{{\rm{A}}_{{\rm{i^*}}}}}$, from the BS to ${{{\rm{B}}_{{\rm{i^*}}}}}$, and from ${{{\rm{A}}_{{\rm{i^*}}}}}$ to ${{{\rm{B}}_{{\rm{i^*}}}}}$, respectively.

Since ${R_{{D_{\rm{C}}}}} \gg {R_{{D_{\rm{B}}}}}$, we can approximate the distance as ${d_{{{\rm{A}}_{\rm{i^*}}}}} \approx {d_{{{\rm{C}}_{\rm{i^*}}}}}$. Using a similar approximation method as that used to obtain \eqref{Theta 1}, we calculate $\Theta _1^*$ as follows:
\begin{align}\label{theta1 nearest 1}
&\Theta _1^* = \nonumber\\
&- \int_0^{{\varepsilon _{{{\rm{A}}_i}}}} \chi  {\rm{ }}\int_{{R_{{D_C}}}}^{{R_{{D_{\rm{A}}}}}} {\frac{{{{\left( {1 + r_{\rm{A}}^\alpha } \right)}^2}}}{{\eta \rho }}{e^{ - \left( {1 + r_{\rm{A}}^\alpha } \right)x}}{\Phi ^*}{f_{{d_{{{\rm{A}}_{\rm{i}}}*}}}}\left( {{r_{\rm{A}}}} \right)d{r_{\rm{A}}}dx} ,
\end{align}
where \\ ${\Phi ^*} = \int_0^{{R_{{D_{\rm{B}}}}}} {\left( {1 + r_{\rm{B}}^\alpha } \right)\left( {\ln \chi \frac{{\left( {1 + r_{\rm{B}}^\alpha } \right)\left( {1 + r_{\rm{A}}^\alpha } \right)}}{{\eta \rho }} + 2{c_0}} \right){f_{{d_{{{\rm{B}}_{\rm{i}}}*}}}}\left( {{r_{\rm{B}}}} \right)d{r_{\rm{B}}}} $ and ${f_{{d_{{{\rm{A}}_{\rm{i}}}*}}}}$ is the PDF of the nearest ${{{\rm{A}}_{{\rm{i^*}}}}}$.

Similar to \eqref{PDF Nearest B} and applying stochastic geometry within the ring ${D_{\rm{A}}}$, we obtain ${f_{{d_{{{\rm{A}}_{\rm{i}}}*}}}}\left( {{r_{\rm{A}}}} \right)$ as follows:
\begin{align}\label{PDF Nearest A}
{f_{{d_{{{\rm{A}}_{\rm{i}}}*}}}}\left( {{r_{\rm{A}}}} \right) = {\xi _{\rm{A}}}{r_{\rm{A}}}{e^{ - \pi {\lambda _{{\Phi _{\rm{A}}}}}\left( {r_{\rm{A}}^2 - R_{{D_{\rm{C}}}}^2} \right)}},
\end{align}
where ${\xi _{\rm{A}}} = \frac{{2\pi {\lambda _{{\Phi _{\rm{A}}}}}}}{{1 - {e^{ - \pi {\lambda _{{\Phi _{\rm{A}}}}}\left( {R_{{D_{\rm{A}}}}^2 - R_{{D_{\rm{C}}}}^2} \right)}}}}$.

Substituting \eqref{PDF Nearest A} and \eqref{PDF Nearest B} into \eqref{theta1 nearest 1}, and using the Gaussian-Chebyshev quadrature approximation, ${\Phi ^*}$ can be expressed as follows:
\begin{align}\label{Phi nearest}
{\Phi ^*} \approx& \frac{{{\xi _{\rm{B}}}{\omega _N}{R_{{D_{\rm{B}}}}}}}{2}\sum\limits_{n = 1}^N {\sqrt {1 - {\phi _n}^2} \left( {1 + c_{n*}^\alpha } \right)} \nonumber\\
&\times\left( {\ln {m_{{\rm{B*}}}}\left( {1 + c_{n*}^\alpha } \right) + 2{c_0}} \right){c_{n*}}{e^{ - \pi {\lambda _{{\Phi _{\rm{B}}}}}c_{n*}^2}}.
\end{align}
where  ${m_{{\rm{B*}}}} = \frac{{\chi \left( {1 + r_{\rm{A}}^\alpha } \right)}}{{\eta \rho }}$.

Substituting \eqref{Phi nearest} into \eqref{theta1 nearest 1}, we obtain
\begin{align}\label{theta1 nearest 2}
\Theta _1^* =&  - \frac{{{\xi _{\rm{B}}}{\xi _{\rm{A}}}{\omega _N}{R_{{D_{\rm{B}}}}}}}{{2\eta \rho }}\int_0^{{\varepsilon _{{{\rm{A}}_i}}}} \chi  {e^{ - \left( {1 + r_{\rm{A}}^\alpha } \right)x}} \nonumber\\
&\times\sum\limits_{n = 1}^N {\sqrt {1 - {\phi _n}^2} \left( {1 + c_{n*}^\alpha } \right){c_{n*}}{e^{ - \pi {\lambda _{{\Phi _{\rm{B}}}}}c_{n*}^2}}} {\Delta ^*}dx,
\end{align}
where ${\Delta ^*} = \int_{{R_{{D_C}}}}^{{R_{{D_{\rm{A}}}}}} {\left( {\ln \chi \frac{{\left( {1 + r_{\rm{A}}^\alpha } \right)}}{{\eta \rho }}\left( {1 + c_{n*}^\alpha } \right) + 2{c_0}} \right)} \\\times{\left( {1 + r_{\rm{A}}^\alpha } \right)^2}{r_{\rm{A}}}{e^{ - \pi {\lambda _{{\Phi _{\rm{A}}}}}\left( {r_{\rm{A}}^2 - R_{{D_{\rm{C}}}}^2} \right)}}d{r_{\rm{A}}}$.

Applying Gaussian-Chebyshev quadrature approximation to ${{\Delta ^*}}$, we obtain
\begin{align}\label{delta nearest}
&{\Delta ^*} \approx \frac{{{\omega _K}\left( {{R_{{D_{\rm{A}}}}} - {R_{{D_{\rm{C}}}}}} \right)}}{2}\sum\limits_{k = 1}^K {\sqrt {1 - \psi _k^2} {{\left( {1 + s_k^\alpha } \right)}^2}} \nonumber\\
&\times\left( {\ln \frac{{\chi \left( {1 + s_k^\alpha } \right)\left( {1 + c_{n*}^\alpha } \right)}}{{\eta \rho }} + 2{c_0}} \right){s_k}{e^{ - \pi {\lambda _{{\Phi _{\rm{A}}}}}\left( {s_k^2 - R_{{D_{\rm{C}}}}^2} \right)}}.
\end{align}

Substituting \eqref{delta nearest} into \eqref{theta1 nearest 1}, we obtain
\begin{align}\label{theta1 nearest 2}
&\Theta _1^* = {b_5}\sum\limits_{n = 1}^N {\sqrt {1 - {\phi _n}^2} \left( {1 + c_{n*}^\alpha } \right)} {c_{n*}}{e^{ - \pi {\lambda _{{\Phi _{\rm{B}}}}}c_{n*}^2}}\nonumber\\
&\times\sum\limits_{k = 1}^K {\sqrt {1 - \psi _k^2} {{\left( {1 + s_k^\alpha } \right)}^2}{s_k}{e^{ - \pi {\lambda _{{\Phi _{\rm{A}}}}}\left( {s_k^2 - R_{{D_{\rm{C}}}}^2} \right)}}} \nonumber\\
&\times\underbrace {\int_0^{{\varepsilon _{{{\rm{A}}_i}}}} {\chi {e^{ - \left( {1 + r_{\rm{A}}^\alpha } \right)x}}\left( {\ln \frac{{\chi \left( {1 + s_k^\alpha } \right)\left( {1 + c_{n*}^\alpha } \right)}}{{\eta \rho }} + 2{c_0}} \right)dx} }_{{\Psi ^*}},
\end{align}
where ${b_5} =  - \frac{{{\xi _{\rm{B}}}{\xi _{\rm{A}}}{\omega _N}{\omega _K}{R_{{D_{\rm{B}}}}}\left( {{R_{{D_{\rm{A}}}}} - {R_{{D_{\rm{C}}}}}} \right)}}{{4\eta \rho }}$.

Applying Gaussian-Chebyshev quadrature approximation to ${{\Psi ^*}}$, we obtain
\begin{align}\label{Psi nearest}
{\Psi ^*} \approx &\sum\limits_{m = 1}^M {{\omega _M}\frac{{{\varepsilon _{{{\rm{A}}_i}}}}}{2}\sqrt {1 - \varphi _m^2} } {e^{ - \left( {1 + s_k^\alpha } \right){t_m}}}\nonumber\\
&\times{\chi _{{t_m}}}\left( {\ln \frac{{{\chi _{{t_m}}}\left( {1 + s_k^\alpha } \right)\left( {1 + c_{n*}^\alpha } \right)}}{{\eta \rho }} + 2{c_0}} \right).
\end{align}
Substituting \eqref{Psi nearest} into \eqref{theta1 nearest 2}, we obtain
\begin{align}\label{theta1 nearest 3}
&\Theta _1^* ={\varsigma ^*}\sum\limits_{n = 1}^N {\sqrt {1 - {\phi _n}^2} \left( {1 + c_{n*}^\alpha } \right){c_{n*}}{e^{ - \pi {\lambda _{{\Phi _{\rm{B}}}}}c_{n*}^2}}} \sum\limits_{k = 1}^K {\sqrt {1 - \psi _k^2} }  \nonumber\\
&\times{\left( {1 + s_k^\alpha } \right)^2}{s_k}{e^{ - \pi {\lambda _{{\Phi _{\rm{A}}}}}\left( {s_k^2 - R_{{D_{\rm{C}}}}^2} \right)}}\sum\limits_{m = 1}^M {\sqrt {1 - \varphi _m^2} }\nonumber\\
&\times{e^{ - \left( {1 + s_k^\alpha } \right){t_m}}}{\chi _{{t_m}}}\left( {\ln \frac{{{\chi _{{t_m}}}\left( {1 + s_k^\alpha } \right)\left( {1 + c_{n*}^\alpha } \right)}}{{\eta \rho }} + 2{c_0}} \right).
\end{align}

Conditioned on the number of users in group $\{{{{\rm{A}}_{{\rm{i}}}}}\}$ and $\{{{{\rm{B}}_{{\rm{i}}}}}\}$, we obtain $\Theta _2^*$ as follows:
\begin{align}\label{Theta 2 Nearest A 1}
\Theta _2^* &= \Pr \left( {\left. {{X_{{i^*}}} < {\varepsilon _{{{\rm{A}}_i}}}} \right|{N_{\rm{A}}} \ge 1} \right)\Pr \left( {\left. {{Y_{{i^*}}} < {\varepsilon _{{{\rm{A}}_i}}}} \right|{N_{\rm{B}}} \ge 1} \right) \nonumber\\
&= {F_{X_{i^*}}}\left( {{\varepsilon _{{{\rm{A}}_i}}}} \right){F_{Y_{i^*}}}\left( {{\varepsilon _{{{\rm{A}}_i}}}} \right).
\end{align}

Similar to \eqref{CDF Nearest B 2}, the CDF of ${\rm{A_{i^*}}}$ is given by
\begin{align}\label{CDF Nearest A 1}
&{F_{X_{i^*}}}\left( \varepsilon  \right)\nonumber\\
&={\xi _{\rm{A}}}\int_{{R_{{D_{\rm{C}}}}}}^{{R_{{D_{\rm{A}}}}}} {\left( {1 - {e^{ - \left( {1 + r_{\rm{A}}^\alpha } \right)\varepsilon }}} \right)} {r_{\rm{A}}}{e^{ - \pi {\lambda _{{\Phi _{\rm{A}}}}}\left( {r_{\rm{A}}^2 - R_{{D_{\rm{C}}}}^2} \right)}}d{r_{\rm{A}}}.
\end{align}
Applying the Gaussian-Chebyshev quadrature approximation, we obtain
\begin{align}\label{CDF Nearest A GC}
{F_{X_{i^*}}}\left( \varepsilon  \right) \approx {b_2}\sum\limits_{k = 1}^K {\sqrt {1 - {\psi _k}^2} \left( {1 - {e^{ - (1 + s_k^\alpha )\varepsilon }}} \right){s_k}{e^{ - \pi {\lambda _{{\Phi _{\rm{A}}}}}s_k^2}}}.
\end{align}

Substituting \eqref{CDF Nearest A GC} and \eqref{CDF Nearest B GC} into \eqref{Theta 2 Nearest A 1} and using a high SNR approximation, we obtain
\begin{align}\label{Theta 2 Nearest A 2 appro}
\Theta _2^* \approx& {b_2}{b_3}\sum\limits_{k = 1}^K {\sqrt {1 - {\psi _k}^2} (1 + s_k^\alpha ){s_k}{e^{ - \pi {\lambda _{{\Phi _{\rm{A}}}}}s_k^2}}} \nonumber\\
&\times \sum\limits_{n = 1}^N {\left( {\sqrt {1 - {\phi _n}^2} \left( {1 + c_{n*}^\alpha } \right){c_{n*}}{e^{ - \pi {\lambda _{{\Phi _{\rm{B}}}}}c_{n*}^2}}} \right)}.
\end{align}

Combining \eqref{Theta 2 Nearest A 2 appro} and \eqref{theta1 nearest 2}, we obtain \eqref{OP Nearest A GC}.

The proof is completed.
\section*{Appendix~D: Proof of Theorem~\ref{theorem:5}} \label{Appendix:D}
\renewcommand{\theequation}{D.\arabic{equation}}
\setcounter{equation}{0}
We express the outage probability for ${{{\rm{A}}_{{\rm{i'}}}}}$ by applying ${X_{{i^*}}} \to {X_i}$, ${Y_{{i^*}}} \to {Y_i}$, and ${Z_{{i^*}}} \to {Z_i}$ in \eqref{OP A 1} and obtain
\begin{align}\label{OP nearest B 1}
&{P_{{{\rm{A}}_{{\rm{i'}}}}}} = \underbrace {\Pr \left( {\left. {\frac{{\rho {X_{i'}}{{\left| {{p_{i1}}} \right|}^2}}}{{\rho {X_{i'}}{{\left| {{p_{i2}}} \right|}^2} + 1}} < {\tau _1},\frac{{\rho {Y_{{i^*}}}{{\left| {{p_{i1}}} \right|}^2}}}{{\rho {{\left| {{p_{i2}}} \right|}^2}{Y_{{i^*}}} + 1}} < {\tau _1}} \right|{\rm V}} \right)}_{{\Theta _2}^\prime }\nonumber\\
&+  \underbrace {\Pr \left( {\left. {{Z_{i'}} < \frac{{{\tau _1} - \frac{{\rho {X_{i'}}{{\left| {{p_{i1}}} \right|}^2}}}{{\rho {X_{i'}}{{\left| {{p_{i2}}} \right|}^2} + 1}}}}{{\eta \rho \left( {{Y_{{i^*}}} - {\varepsilon _{{{\rm{A}}_i}}}} \right)}},{X_{i'}} < {\varepsilon _{{{\rm{A}}_i}}},{Y_{{i^*}}} > {\varepsilon _{{{\rm{A}}_i}}}} \right|{\rm V}} \right)}_{{\Theta _1}^\prime },
\end{align}
where ${X_{i'}} = \frac{{{{\left| {{h_{{{\rm{A}}_{\rm{i}}}}}} \right|}^2}}}{{1 + d_{{{\rm{A}}_{{\rm{i'}}}}}^\alpha }}$ and ${Z_{i'}} = \frac{{{{\left| {{g_i}} \right|}^2}}}{{1 + d_{{{\rm{C}}_{{\rm{i'}}}}}^\alpha }}$. Here ${d_{{{\rm{A}}_{{\rm{i'}}}}}}$ and ${d_{{{\rm{C}}_{{\rm{i'}}}}}}$ are distances from the BS to ${{{\rm{A}}_{{\rm{i'}}}}}$ and from ${{{\rm{A}}_{{\rm{i'}}}}}$ to ${{{\rm{B}}_{{\rm{i^*}}}}}$, respectively.

Since ${R_{{D_{\rm{C}}}}} \gg {R_{{D_{\rm{B}}}}}$, we can approximate the distance as ${d_{{{\rm{A}}_{\rm{i'}}}}} \approx {d_{{{\rm{C}}_{\rm{i'}}}}}$. Using a similar approximation method as that used to get \eqref{Theta 1}, we first calculate ${{\Theta _1}'}$ as follows:
\begin{align}\label{theta2 farthest 1}
{{\Theta _1}'} =& - \int_0^{{\varepsilon _{{{\rm{A}}_i}}}} \chi  \int_{{R_{{D_C}}}}^{{R_{{D_{\rm{A}}}}}} {\frac{{{{\left( {1 + r_{\rm{A}}^\alpha } \right)}^2}}}{{\eta \rho }}{e^{ - \left( {1 + r_{\rm{A}}^\alpha } \right)x}}} \nonumber\\
&\times\int_0^{{R_{{D_{\rm{B}}}}}} {\left( {1 + r_{\rm{B}}^\alpha } \right)\left( {\ln \chi \frac{{\left( {1 + r_{\rm{B}}^\alpha } \right)\left( {1 + r_{\rm{A}}^\alpha } \right)}}{{\eta \rho }} + 2{c_0}} \right)} \nonumber\\
&\times {f_{{d_{{{\rm{B}}_{\rm{i}}}*}}}}\left( {{r_{\rm{B}}}} \right)d{r_{\rm{B}}}{f_{{d_{{{\rm{A}}_{{\rm{i'}}}}}}}}\left( {{r_{\rm{A}}}} \right)d{r_{\rm{A}}}dx,
\end{align}
where ${f_{{d_{{{\rm{A}}_{\rm{i'}}}}}}}\left( {{r_{\rm{A}}}} \right)$ is the PDF for the farthest ${{{\rm{A}}_{{\rm{i'}}}}}$.

Similar to \eqref{PDF Nearest B} and applying stochastic geometry within the ring ${D_{\rm{A}}}$, we can obtain ${f_{{d_{{{\rm{A}}_{\rm{i'}}}}}}}\left( {{r_{\rm{A}}}} \right)$ as follows:
\begin{align}\label{PDF farthest A}
{f_{{d_{{{\rm{A}}_{\rm{i}}}'}}}}\left( {{r_{\rm{A}}}} \right) = {\xi _{\rm{A}}}{r_{\rm{A}}}{e^{ - \pi {\lambda _{{\Phi _{\rm{A}}}}}\left( {R_{{D_{\rm{A}}}}^2 - r_{\rm{A}}^2} \right)}}.
\end{align}

Conditioned on the number of ${\rm{A_{i'}}}$ and ${\rm{B_{i^*}}}$, we obtain
\begin{align}\label{Theta 2 farthest A 1}
{\Theta _2}' &= \Pr \left( {\left. {{X_{i'}} < {\varepsilon _{{{\rm{A}}_i}}}} \right|{N_{\rm{A}}} \ge 1} \right)\Pr \left( {\left. {{Y_{{i^*}}} < {\varepsilon _{{{\rm{A}}_i}}}} \right|{N_{\rm{B}}} \ge 1} \right)\nonumber\\
&= {F_{{X_{i'}}}}\left( {{\varepsilon _{{{\rm{A}}_i}}}} \right){F_{{Y_{{i^*}}}}}\left( {{\varepsilon _{{{\rm{A}}_i}}}} \right).
\end{align}

Following a similar procedure as that used to obtain $\Theta _1^*$ and $\Theta _2^*$ in Appendix~B, we can obtain ${{\Theta _1}'}$ and ${{\Theta _2}'}$. Then combining  ${{\Theta _1}'}$ and ${{\Theta _2}'}$, the general case \eqref{OP farthest A GC} is obtained. For the special case $\alpha=2$, following a method similar to that used to calculate \eqref{OP Nearest A GC alpha 2}, we can obtain \eqref{OP farthest A GC alpha 2}.

The proof is completed.

\bibliographystyle{IEEEtran}
\bibliography{mybib}

\end{document}